\documentclass[fleqn,usenatbib]{mnras}

\usepackage[T1]{fontenc}
\usepackage{ae,aecompl}

\usepackage{graphicx}	
\usepackage{amsmath}	
\usepackage{amssymb}	
\usepackage{threeparttable}
\usepackage{color}

\usepackage{ulem}

\usepackage{newtxtext,newtxmath}

\title[CNN for PPDs]{PGNets: Planet mass prediction using convolutional neural networks for radio continuum observations of protoplanetary disks}

\author[S. Zhang et al.]{
Shangjia Zhang$^{1}$ \thanks{E-mail: shangjia.zhang@unlv.edu}, 
Zhaohuan Zhu$^{1}$,
Mingon Kang$^{2}$
\\
$^{1}$Department of Physics and Astronomy, University of Nevada, Las Vegas, 4505 S. Maryland Pkwy, Las Vegas, NV, 89154, USA\\
$^{2}$Department of Computer Science, University of Nevada, Las Vegas, 4505 S. Maryland Pkwy, Las Vegas, NV, 89154, USA\\
}

\pubyear{2021}

\begin{document}
\label{firstpage}
\pagerange{\pageref{firstpage}--\pageref{lastpage}}
\maketitle

\begin{abstract}
We developed Convolutional Neural Networks (CNNs) to rapidly and directly infer the planet mass from radio dust continuum images. 
Substructures induced by young planets in protoplanetary disks can be used to infer the potential young planets' properties. Hydrodynamical simulations have been used to study the relationships between the planet's properties and these disk features. However, these attempts either fine-tuned numerical simulations to fit one protoplanetary disk at a time,
which was time-consuming, or azimuthally averaged simulation results to derive some linear relationships between the gap width/depth and the planet mass, which lost information on asymmetric features in disks. To cope with these disadvantages, we developed Planet Gap neural Networks (PGNets) to infer the planet mass from 2D images. We first fit the gridded data in Zhang et al. (2018) as a classification problem. Then, we quadrupled the data set by running additional simulations with near-randomly sampled parameters, and derived the planet mass and disk viscosity together as a regression problem. The classification approach can reach an accuracy of 92\%, whereas the regression approach can reach 1$\sigma$ as 0.16 dex for planet mass and 0.23 dex for disk viscosity. We can reproduce the degeneracy scaling $\alpha$ $\propto$ $M_p^3$ found in the linear fitting method, which means that the CNN method can even be used to find degeneracy relationship. The gradient-weighted class activation mapping effectively confirms that PGNets use proper disk features to constrain the planet mass. We provide programs for PGNets and the traditional fitting method from Zhang et al. (2018), and discuss each method's advantages and disadvantages.

\end{abstract}





\section{Introduction}\label{sec:intro}
Detecting young planets in protoplanetary disks is essential to infer where and when planets form and how massive they are, putting stringent constraints on planet formation theory. Unfortunately, despite thousands of exoplanets having been discovered, only a few of them are around young stars within 10 million years old. There are even fewer young forming planets found in dusty protoplanetary disks. One notable example is PDS 70 system, where two young planets with several Jupiter mass have been discovered within the dusty cavity of 80 au \citep{Keppler2018,Muller2018,Wagner2018,Haffert2019,Isella2019,Christiaens2019,Wang2020,Hashimoto2020}. However, such firm detection seems to be rare (e.g., \citealt{Zurlo2020}), and we are not expecting to directly detect a planet whose mass is less than one Jupiter mass \citep{Ruane2017}. There are other promising methods to detect young planets in disks, such as using disk kinematic features influenced by the planet \citep{Perez2015b,teague18a,pinte2018,Rabago2021,Izquierdo2021}. However, all these methods can only detect planets that are more massive than Jupiter. The only method that can detect planets less massive than Jupiter is to use gaps in the dust continuum images (e.g., \citealt{zhang18}). Since dust particles drift to the local pressure maximum, even small gas perturbations by a low mass planet can lead to observable dust gaps \citep{paardekooper06,zhu14,rosotti2016,dipierro2017,dong18b}.

Hydrodynamical simulations with dust particles have been carried out to infer the planet properties from radio observations. At early times when ALMA high resolution observations were scarce, numerical simulations were fine-tuned to fit one source at a time (e.g. \citealt{dong2015a,Dipierro2015a, picogna2015, dipierro2018}). However, when high angular-resolution 
surveys became available (e.g., \citealt{andrews18b,long18,Cieza2021}), it was impractical to run direct numerical simulations for each source. Thus, relationships between the gap and planet properties have been solved. For gaseous gaps, such relationships have been well characterized \citep{fung14,kanagawa_depth,kanagawa_width}. However, the relationships are less clear for dusty gaps mainly because the gap width/depth can vary significantly with different sized particles in disks. \cite{Lodato2019} assumed that the dusty gap width is scaled with the planet Hill radius to derive the masses of planets in the Taurus survey \citep{long18}.
\cite{rosotti2016} carried out simulations to derive the relationship between the planet mass and the distance between the planet and the pressure maximum at the outer gap edge. However, \cite{rosotti2016} did not consider the effects of particle size and disk viscosity, both of which can change the gap-planet relationship significantly. A detailed study was done by
\cite{zhang18}, who carried out a large grid of planet-disk interaction simulations with dust particles, generated synthetic observations, and derived relationships between the planet mass, the disk scale height, the disk viscosity, and the particle size in disks\footnote{We use ``the linear fitting method'' to denote this method throughout this paper.}. Such relationships have been used to derive the young planet population from the DSHARP survey \citep{andrews18b, zhang18}. 

However, the approach in \cite{zhang18} still suffers several shortcomings. First, the procedures to derive the planet mass are relatively complicated.
Disk viscosity, scale height, and particle size need to be constrained by other methods and specified beforehand. Based on these parameters, different fitting formulae need to be adopted. To make the method in \citet{zhang18} easier to use, parallel to the method presented in this paper, 
we provide new {\sc Python} programs for the \cite{zhang18} linear fitting method\footnote{The code is available at \url{https://github.com/zhangsj96/DSHARPVII.git}.}, which can automatically find the planet mass after the parameters are specified. The second shortcoming, which is intrinsic to the method itself, is that the synthetic images generated from simulations were azimuthally averaged before deriving the relationship between the planet mass and gap widths/depths. Thus, all information from non-axisymmetric features was lost. These features have rich information on disk and planet properties. For example, an eccentric gap suggests a massive planet. A lopsided disk with a large 
intensity asymmetry indicates a low viscosity or large particles.

To directly extract information from 2D images, in this work, we adopted machine learning techniques.
Machine learning techniques have been widely used across the astronomical community for decades, and here we briefly describe several works related to computer vision image tasks, using Convolutional Neural Networks (CNNs), specifically.

CNNs have been used widely in galactic and extra-galactic studies.
\citet{dieleman15} is one of the first works that use modern CNN networks. It originated from an international competition \textit{Galaxy Challenge} that aimed to build automated tools for galaxy morphology classification based on annotated images from the Galaxy Zoo project. They applied CNNs to the morphological classification of crowd-sourcing annotated images and achieved $>$ 99\% accuracy, which would benefit analysis in future large galaxy surveys such as Vera C. Rubin Observatory. 
\citet{hezaveh17} used CNNs to automate analysis of strong gravitational lenses. The traditional method with maximum likelihood modeling requires human expertise and is time-consuming. The network can quantify image distortions caused by strong gravitational lensing and estimate these structures' corresponding matter distribution with comparable accuracy as the traditional method, but ten million times faster. Non-experts can quickly obtain lensing parameters for a large sample of data.
\citet{hassan19} utilized CNNs to identify reionization sources from 21 cm maps. Active galactic nuclei and star-forming galaxies are two leading sources that reionized our Universe. CNNs were trained to distinguish the sources on the 21 cm images. The technique would aid power-spectrum observations and provide extra information to break degeneracies between a broad range of reionization models. The classification accuracy is between 92-100\%, depending on the redshift and neutral fraction range.

CNNs have also been used in the Solar System study. 
\citet{lieu19} trained CNNs on simulated observations of an upcoming mission \textit{Euclid} for Solar system objects identification. They used transfer learning (i.e., training on several established CNN architectures with some modifications) on a relatively small data set. Their best model correctly identified objects with a top accuracy of 94\%, successfully separating solar system objects from other astronomical sources.

In the field of star formation, identifying signatures of stellar feedback in molecular clouds used to mainly rely on visual inspection. \citet{vanoort19} ran 3D MHD simulations with stellar feedback and produced 2D synthetic CO continuum images. They trained CNNs on synthetic data and identified shells in real observations. Later they extended the work to 3D so that they could make full use of molecular line spectra datacube. They found stellar feedback bubbles and predicted feedback properties \citep{xu20a}, and identified 20 new outflows \citep{xu20b} that were missed by previous visual inspections.

While CNNs have not been applied to the planet-disk interaction study, general machine learning techniques are receiving more attention in the field. Recently \citet{auddy20} used fully connected neural networks to fit the relationship between the planet mass and parameters such as gap width, aspect ratio, viscosity, dust-to-gas ratio, Stokes number, and density profile. Compared to previous fitting methods, the work is the first to fit the relationship non-linearly. Since a deep neural network is good at fitting problems that are intrinsically non-linear, their estimated planet mass follows closer to the simulation data given a multi-dimensional input. Nevertheless, users still need to provide inputs that are barely constrained from observations, and asymmetric information is still lost when 2D images are converted to 1D radial profiles. While we were modifying this paper after the first referee report, \cite{Auddy2021} published a CNN approach for the 2D images, which alleviated some of these shortcomings. 

Our aim in this paper is to infer the planet mass from the 2D observational images directly \footnote{The code is available at \url{https://github.com/zhangsj96/PGNets.git}.}. In Section \ref{sec:background} we briefly introduce the background and the basic glossary of CNNs. In section \ref{sec:method} we describe the simulation setup, synthetic observation production, prepossessing, augmentation, and the neural network setup and training. In Section \ref{sec:results} we analyze the results, apply the networks to several gaps in DSHARP observations, and compare the derived planet masses to those from the previous method. After a short discussion in Section \ref{sec:discussion}, we conclude our paper in Section \ref{sec:conclusion}.

\section{Background} \label{sec:background}

\subsection{Regular neural network}
The most common subset of deep learning is regular neural networks (or fully-connected neural networks). Neural networks receive a vector and transform it through a series of hidden layers. Each hidden layer is made up of a set of neurons, where each neuron is fully connected to all neurons in the previous layer. The last fully-connected layer is called the ``output layer''. For classification problems, it gives scores for different classes. For regression problems, it predicts continuous values.

Each neuron has some parameters to be tuned, which can be accomplished by training the model. The training is a process of minimizing the loss function and updating parameters through back-propagation. The data are separated as training, validation and testing sets. The training data are used to feed into the neural network. The validation data are not used in training the model but are used as a metric to monitor the training result at every epoch. The testing data are used to evaluate the model accuracy after the training is completed.

\subsection{Convolutional neural network (CNN)}
For regular neural networks, the input vector could be extremely large if the input is an image that is represented by either a flattened matrix (e.g., a gray-scale image) or a flattened tensor (e.g., an RGB-colored image). There are correlations between neighboring pixels and different color channels in an image, but a vector representation loses such correlations. Thus, regular neural networks are not ideal for training image data. The convolutional operation naturally takes the local connections into account. To that end, CNNs are powerful in fitting image data. As a variation of regular neural networks, their primary unit of computation is the convolutional operation instead of simple matrix multiplication. A layer of a convolutional network has neurons arranged in three dimensions: width, height, and depth. A convolutional kernel will be operated on this 3D tensor, and the output becomes the next layer. Usually, the network's width and height become smaller for later layers, while the depth becomes deeper. The process of downsampling the feature map is called pooling. 
\citet{lecun98} introduced LeNet to recognize hand-written digit characters. It reached a very high performance and brought artificial neural networks into popularity.

\subsection{Residual neural network}
A residual neural network (ResNet) is a kind of CNNs that has connections even between skipping (non-neighboring) layers \citep{he16a}. It has shortcuts to jump over some layers. Typical ResNet models are implemented with double or triple skips that contain ReLU and batch normalization \citep{ioffe15} in between. Skipping effectively simplifies the networks and reduces the parameters. It also avoids the problem of vanishing gradients so that the network can go deeper than traditional CNNs while still improving the performance.

\section{Method} \label{sec:method}

\begin{figure*}
\includegraphics[width=\linewidth]{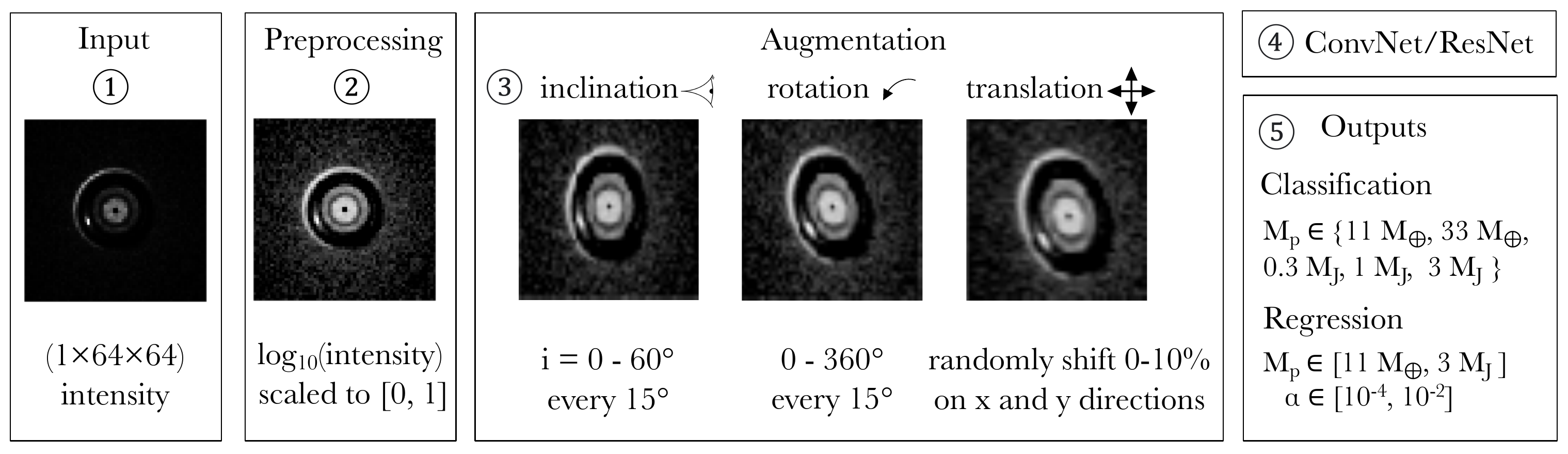}
\caption{A schematic view of our work from input to output. There are five steps. First, we prepared the input from simulations or observations. Second, the image was prepossessed and normalized. Third, the image was augmented for different inclinations, rotations, and translational shifts. Then it flowed into the neural network. Finally, we obtained class scores and chose the label with the highest score for the classification problem. A planet mass, viscosity pair was returned for the regression problem.
\label{fig:schematic}} 
\end{figure*}

A schematic view of our method from input to output is summarized in Figure \ref{fig:schematic}. We first introduce how we convert the simulations to synthetic observations (Section \ref{sec:simulation}), and then discuss the prepossessing (Section \ref{sec:preprocessing}) and augmentation (Section \ref{sec:augmentation}) steps to make the network robust. We layout the PGNet structures (VGG-like or ResNet classification; Section \ref{sec:setup}; regression; Section \ref{sec:setupreg}) and finally obtain the output prediction.

At the output layer, we tried both classification and regression problems. At first, we treated the fitting of the planet mass as a classification problem, since the data set of \citet{zhang18} are too sparse (only five discrete planet masses) to return continuous-valued predictions. They were used to demonstrate that CNNs can successfully predict planet masses on discrete grids. Then we ran additional 150 simulations to provide more sampling between these grids, and built regression model which predict continuous planet mass and disk viscosity at the same time. The samples were drawn using the Latin hypercube sampling (LHS) \citep{McKay79}.

\subsection{Simulations\label{sec:simulation}}
For the classification problem, we used the results of the planet-disk interaction simulations in \cite{zhang18} on the grddied parameter space (5 planet masses, $M_p$, 3 aspect ratios, $h/r$ , and 3 disk viscosities, $\alpha$). We denote a model with a \{$M_p$, $h/r$, $\alpha$\} pair as a generic model, as this model can be used to generate models with different surface densities and maximum particle sizes. For the regression problem, we added additional 150 simulations with near-randomly generated $M_p$, $h/r$, and $\alpha$. We briefly summarize the simulations here. The simulations were carried out with 2D hydrodynamic code Fargo-ADSG \citep{baruteau2008a, baruteau2008b, baruteau2016}. Dust grains were represented by 200,000 super-particles with different sizes. The Stokes number ($St$) of the particles at $r_p$  ranged from 1.57 $\times 10^{-5}$ to 1.57. The simulations in the gridded parameter space covered three disk viscosities $\alpha$=$10^{-4}$, $10^{-3}$ and $10^{-2}$, three disk aspect ratios at $r_p$ with $h/r$ = 0.05, 0.07 and 0.1, and five planet masses with the planet-star mass ratios ($q$) of 3.3$\times$10$^{-5}$, 10$^{-4}$, 3.3$\times$10$^{-4}$, 10$^{-3}$, 3.3$\times$10$^{-3}$ (which are equivalent to the planet masses of
11 $M_\oplus$, 33 $M_\oplus$, 0.3 $M_J$, 1 $M_J$ and 3 $M_J$ if the central star is a solar mass star). The parameters for the LHS were drawn from  $\alpha$ $\in$ [$10^{-4}$, $10^{-2}$], $h/r$ $\in$ [0.05, 0.1], and  $M_p$ $\in$ [11 $M_\oplus$, 3 $M_J$]. They were near-uniformly drawn in the interval of $h/r$, log($\alpha$) and log($M_p$).\footnote{We used \texttt{pyDOE} (\url{https://github.com/tisimst/pyDOE}) to generate parameters. The generated parameters can be found at \url{https://github.com/zhangsj96/PGNets.git}.} We initialized the gas surface density as
\begin{equation}
\Sigma_g(r)=\Sigma_{g,0}(r/r_{0})^{-1}\,,
\end{equation}
where $r_{0}$  is the position of the planet and we set $r_0=r_p=1$. Our numerical grid extended from 0.1 $r_0$ to 10 $r_0$ in the radial direction and 0 to 2$\pi$ in the $\theta$ direction. The data sets used in this paper are at 1000 planetary orbits.
We assumed locally isothermal equation of state. The temperature at radius $r$ follows $T(r)=T_{0}(r/r_{0})^{-1/2}$.

To convert super-particle distributions to optical depth maps, we used a subset of particles and gave them different weights depending on their opacity, sizes, and surface density at their locations. We adopted DSHARP opacity \citep{birnstiel18} and neglected scattering. These particles were interpolated onto 1200 x 1200 regular grid with physical dimensions as $10$ $r_p$ $\times$ $10$ $r_p$. Then we smoothed the gridded data with a Gaussian kernel similar to the resolution of ALMA. Gaussian $\sigma$ = 2 pixels, which is 0.03 $r_p$. If the planet is at 20 au, this is the same as the resolution of a circular beam with FWHM = 1.4 au.
Finally, we calculated the brightness temperature or intensity for each grid cell as
\begin{equation}
    T_b(x,y) = T_d(r)( 1-e^{-\tau(x,y)})\,,
    \label{eq:normTb}
\end{equation}
and we assumed that the midplane temperature follows
\begin{equation}
    T_d(r)=T_d(r_0)\left(\frac{r}{r_0}\right)^{-0.5}\,.
\end{equation}
For more details please see \citet{zhang18}.

In principle, the real synthetic image should be calculated with assumptions of detailed observational setups (e.g., antenna array configuration and integration time) and CLEAN methods. However, this is unrealistic since (a) they depend on specific observational setups and data reduction methods, which cannot be covered thoroughly, and (b) these time-consuming steps need to be applied on each augmented image (Section \ref{sec:augmentation}), which cannot be realized within our computational power. Instead, in Section \ref{sec:CASA} we will use one case to demonstrate that neglecting these steps does not affect the correctness of the prediction.

\begin{figure*}
\includegraphics[width=\linewidth]{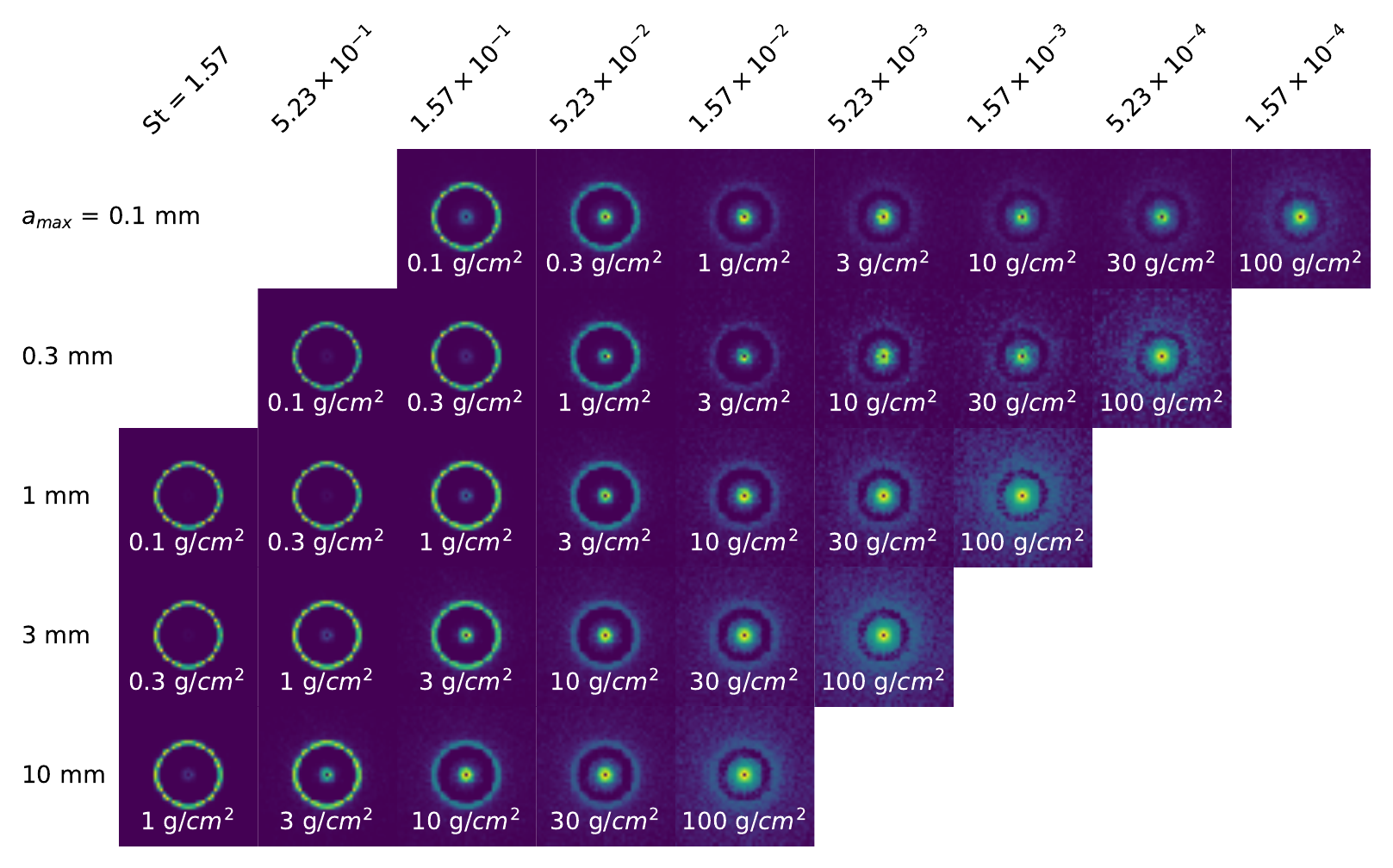}  
\caption{\label{fig:intensity} 1.3 mm dust continuum intensity maps for the M$_p$ = 1 M$_J$, $\alpha$= $10^{-3}$ and $h/r$=0.07 case with different $a_{max}$ and $\Sigma_{g,0}$. In each row, the $a_{max}$ is the same with increasing surface density from left to right. The Stokes number at the planet's location is the same at each column and decreasing from left to right.}
\end{figure*}

To generate synthetic radio continuum images of disks having various dust size distributions and disk surface densities, we chose 5 different maximum particle sizes $a_{max}$ = 0.1 mm, 0.3 mm, 1 mm, 3 mm and 1 cm, and 7 different gas surface densities $\Sigma_{g, 0}$ = 0.1 $\mathrm{g/cm^2}$, 0.3 $\mathrm{g/cm^2}$, 1 $\mathrm{g/cm^2}$, 3 $\mathrm{g/cm^2}$, 10 $\mathrm{g/cm^2}$, 30 $\mathrm{g/cm^2}$ and 100 $\mathrm{g/cm^2}$. These combinations of $a_{max}$ and $\Sigma_{g, 0}$ correspond to 9 different characteristic Stokes numbers (St $\propto$ $a_{max}$/$\Sigma_{g, 0}$). Figure \ref{fig:intensity} shows dust continuum intensity at 1.3 mm for a case with $M_p$ = $M_J$, $\alpha$= $10^{-3}$ and $h/r$=0.07, with all combinations of $a_{max}$ and $\Sigma_{g, 0}$. Note that for $a_{max}$ = 3 mm (and 1 cm), the lowest (two) surface density case(s) exceed the upper limit of the particles' Stokes number in our simulation. In total, different $a_{max}$ and $\Sigma_{g,0}$ lead to 32 combinations (7+7+7+6+5) for each generic model. Since no dust growth and back-reaction are included,  the dust drift velocity only depends on the Stokes number. This is why the gaps look similar for a given Stokes number. They are still different in that (a) the opacity is dependent on the maximum particle size and (b) the dust surface density is different. Thus, the radial profile of the optical depth is different.  When the disk changes from the optically thin to thick regime (from top to bottom panels), the intensity maps become smoother. 

We explored different dust size distributions by choosing three different power-law indices ($p$; n(a)$\propto$ $a^{-p}$) being 3.5, 3 and 2, but their intensity maps are very similar. This could lead to some duplication between training and testing sets. For this reason, we only used one power-law index $p$ = 3.5.

An image with too many pixels takes a large amount of memory, does not help the training, and even leads to overfitting. Thus, we downgraded the image size and found that a 64 $\times$ 64 image is good enough to preserve features and help the training. Note that the real observation image will also be fed into the network using this resolution. To convert an observation to the input format, a simple linear interpolation suffices. For our models, we selected the central squared 3$r_p$ $\times$ 3$r_p$ region. Outside of this region, there is little emission from our simulations.  The data are 2D (gray-color, 1$\times$64$\times$64) images of disk emission. There are 5 $\times$ 3  $\times$ 3  $\times$ 32 = 1440 models for the classification problem and (45+150) $\times$ 32 = 6240 models for the regression problem.

\subsection{Preprocessing \label{sec:preprocessing}}

We found that adding some noise to images helped the training, so we added 10\% RMS noise to each image. The initial value of the intensity in the image can be different by orders of magnitude. In neural networks, it is essential to normalize the input data.  We found that taking the logarithm of the value works best for the training. We scaled the image values from 0 to 1. We also set an emission floor as 1\% of the maximum emission intensity, so that values below that became zero. In expression,

\begin{align}
\text{pixel value} = \, \begin{cases}
\frac{\mathrm{log_{10}}(I) - \mathrm{log_{10}}(I_{max}) + 2}{2}
  \ \text{if} \ \mathrm{log_{10}}(I) > \mathrm{log_{10}}(I_{max} - 2) , \\
0 \ \ \ \ \ \ \ \ \ \ \ \ \ \ \ \ \ \ \ \ \ \ \ \ \ \ \ \ \ \ \ \ \ \  \text{elsewhere}.                                            
 \end{cases}
 \label{eq:gaussianbump}
\end{align}
This choice of the floor value can also be justified for observational data since the sensitivity limit for these high-resolution ALMA observations is usually between 1\% to 0.1\%. The gap region has low values, and the ring region has high values. Considering that previous studies used gap shapes to infer the planet mass, we reversed the value as (1-value) so that the gap had high values. It turned out that this procedure had little effect on the results. Since this procedure is well defined, it is also simple to apply it to the real observation and rapidly convert that to the network input. Note that our main focus is on the gap shape (width and depth, and asymmetric features). The normalization process we adopted removed the information of the absolute value of the emission. Ideally, the absolute value of the emission can provide extra information and can be considered in future works.

\subsection{Data augmentation \label{sec:augmentation}}

In real observations, disks are hardly to be exactly face-on (e.g., the configuration in Figure \ref{fig:intensity}). As long as the gap is spatially resolved and the disk is not too edge-on, we should still obtain information from the image. Thus, we inclined the disk from the initial data from 0$^\circ$ to 60$^\circ$, spacing every 15$^\circ$. This was simply done by stretching the y-direction of the optical depth map, $\tau(x,y)$ while fixing the x-direction size. The optical depth is increased accordingly. This can be justified considering that mm dust is highly settled in protoplanetary disks \citep{pinte2016}. With this data augmentation, the observational data for disks with any inclinations can be directly used as an input for our networks.

To explore the rotational symmetry \citep{dieleman15}, we rotated the disk every 15$^\circ$, from 0$^\circ$ to 360$^\circ$ (0$^\circ$ and 360$^\circ$ are only counted once). The degree of rotation is also called position angle.
Note that the rotation and inclination variations were done on the original 1200 $\times$ 1200 (10 $r_p$ $\times$10 $r_p$) data to avoid the sharp edges due to missing values, but whether including this step did not affect the results.

We also explored translational symmetry by randomly shifting the image in x and y directions from 0\% to 10\%. Even though CNNs should conserve translational symmetry by themselves, the fitting result was slightly improved. This is also helpful for the prepossessing of observational data since the disk does not need to be perfectly centered when they are fed into the networks.

Every one of the face-on images can generate thousands of augmented data. However, we still separated the training, validation, and testing sets with those original data. Otherwise, if we separate all the data after the augmentation, the fitting accuracy will actually increase since different augmented data generated from a single model can look very similar. However, this increase of the accuracy is unreal due to the similar data in training, validation, and tests. Thus, we separated the data sets before the augmentation. 

\subsection{Classification models \label{sec:setup} }

We first tested out CNN models as a classification problem on the gridded data set in \citet{zhang18}. We set up the network as a classification problem instead of a regression problem since our origional simulations only have five discrete planet masses. We built two neural networks for comparison, one architecture similar to VGG-16 \citep{simonyan14} and the other adapted from ResNet \citep{he16a}. We chose not to use the original VGG-16 architecture since it introduced more parameters but did not improve the performance.

\begin{figure*}
\includegraphics[width=\linewidth]{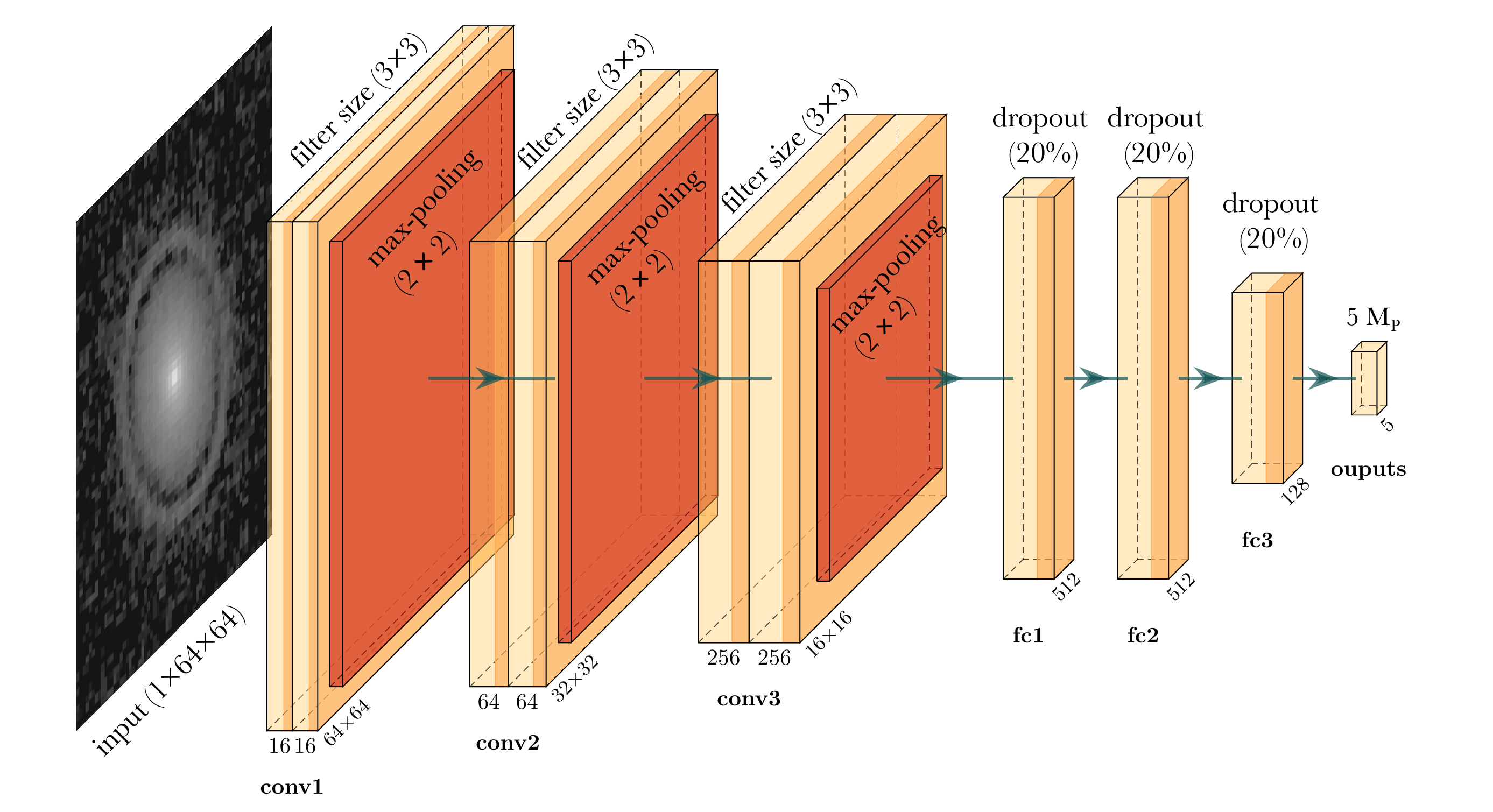}\caption{The architecture of VGG-like network. The image as an input is the log$_{10}$ of the intensity, scaled between zero and one with only one channel. The example image is preprocessed image of GW Lup \citep{andrews18b}. Then there are six convolutional layers and three fully-connected layers.
\label{fig:architecture}} 
\end{figure*}

Figure \ref{fig:architecture} shows the architecture of VGG-like network. As mentioned in the previous section, the input data are gray-color images (1$\times$ 64 $\times$ 64), with values ranging from 0 to 1. Then it follows two convolutional layers with 16 channels (depth=16). 
In this specific neural network, the convolutional filter size is always 3 $\times$ 3 with 1 stride. 
The padding is on image edges so that the tensor keeps the same number of points (64) in both x and y directions. We also used ReLU (Rectification Linear Unit, i.e., max[0, x]), as the activation function for all layers. After a 2 $\times$ 2 max-pooling operation, the image dimension becomes 32 $\times$ 32. Then it follows two convolutional layers with 64 channels and another max-pooling. The last two convolutional layers have 256 channels before a max-pooling. Then it connects to 3 fully-connected layers, two with 512 nodes and one with 128 nodes. These fully-connected layers all have 20\% dropout rates to avoid overfitting. Finally, the output layer has 5 nodes which stand for 5 different planet masses. There are 9,504,565 trainable parameters in total. We found that adding more layers did not increase the accuracy.

The learning rate is a hyper-parameter that controls the step size in an optimization process while it is minimizing the loss function. An analogy of it is the step size in solving differential equations. We adopted the initial learning rate as 0.001. We chose the sparse categorical cross-entropy as the loss function and  \texttt{ADAM} as the optimizer \citep{kingma14}. We defined accuracy of the method as 
\begin{equation}
    \mathrm{accuracy} = \frac{\mathrm{TP+TN}}{\mathrm{TP+TN+FP+FN}},
    \label{eq:accuracy}
\end{equation}
where TP, TN, FP and FN stand for true positive, true negative, false positive and false negative, respectively.


To balance the trade-off between rate of convergence and overshooting, we adopted an adaptive learning rate. The learning rate varying as the training epoch is a cosine function so that the learning rate becomes much smaller at later training stages. The batch size is 128. The training, validation, and testing sets were split as 60\%, 20\%, and 20\% of  the 1440 models that were randomly shuffled. We also included different inclinations and rotations for the testing set, so there are total 1440$\times$0.2$\times$5$\times$24 = 34560 testing images. The training was done on a single NVIDIA GPU GeForce RTX 2080 Ti 12GB. We used \texttt{Tensorflow v2} \citep{tensorflow2015-whitepaper} to build the network. The VGG-like network took less than an hour, while the ResNet model took two hours.

For the residual network, our network was adopted from ResNet v2 \citep{he16b} with three skips. It has a bottleneck layer with stacks of batch normalization, activation (ReLU) and convolutional layers. The batch normalization can make ANN faster and more stable. The first shortcut connection per layer is 1 x 1 convolution and the second and onward shortcut connection is identity. At the beginning of each stage, the feature map size is halved by a convolutional layer with 2 strides, while the number of filter maps is doubled. Within each stage, the layers have the same number of filters and the same filter map sizes. There are 22 convolutional layers in total, but with only 576,357 parameters, an order of magnitude smaller than the VGG-like model. The initial learning rate is 0.005. Other setups are the same as the VGG-like model.

\subsection{Regression \label{sec:setupreg}}

After carrying out additional simulations with the randomly-generated parameters from the LHS, we were able to fit the planet mass in a continuous space. Thus, we fitted the combined data set as a regression problem. We used mean square error (MSE) as the loss function. We made the output layer as a $M_p$-$\alpha$ pair, as MSE can evaluate vectors (e.g., \citealt{alibert19}). In this way, the planet mass and disk viscosity can be predicted together. We made viscosity as an output since a direct measurement of gas turbulence is difficult (e.g., \citealt{flaherty2018}). We adopted the ResNet architecture as mentioned in the previous subsection.

Slightly different from Section \ref{sec:setup}, we split 195 generic models into 175 and 20. The selection was random but can be reproduced in our repository. We then separated the first 175 models, together with their generated models with combinations of surface densities and maximum particle sizes into 60\%, 20\%, 20\% splits as training, validation and test data. Finally, we used the rest 20 generic models as a ``genuine test'', as the input parameters of these models are completely outside the parameter space of \{$M_p$, $h/r$, $\alpha$\} pairs in the training. This additional step can test if the overfitting happens in the \{$M_p$, $h/r$, $\alpha$\} space. This is a more rigorous test than that in Section \ref{sec:setup} since we separate the data even before generating the face-on images with different $a_{max}$ and surface densities. We used callback function \texttt{ReduceLROnPlateau} to reduce learning rate when the loss had stopped improving.
We kept other hyper-parameters the same as what were mentioned in previous subsections.

\section{Results} \label{sec:results}

\subsection{Classification}
The accuracy of VGG-like model reached a plateau after 10 training epochs but still slowly increased up to 40 epochs. The accuracy of the ResNet model reached a plateau after 90 epochs. The ResNet has slightly higher accuracy than the VGG-like model on the validation set. This is also the case for the testing set. The VGG-like model can reach an accuracy of 89\%, and the ResNet model can reach an accuracy of 92\%. Note that the accuracy reported is the micro accuracy (i.e., applying equation \ref{eq:accuracy} for the whole sample, instead of calculating accuracies for each class and averaging them, which is the macro accuracy). Even though each class (planet mass) has the same amount of sample in the whole data set, in each subsets (training, validation and testing sets), the data are slightly imbalanced due to the shuffling process to separate them. Nevertheless, macro and micro metrics are similar here. 

\begin{figure}
\includegraphics[width=\linewidth]{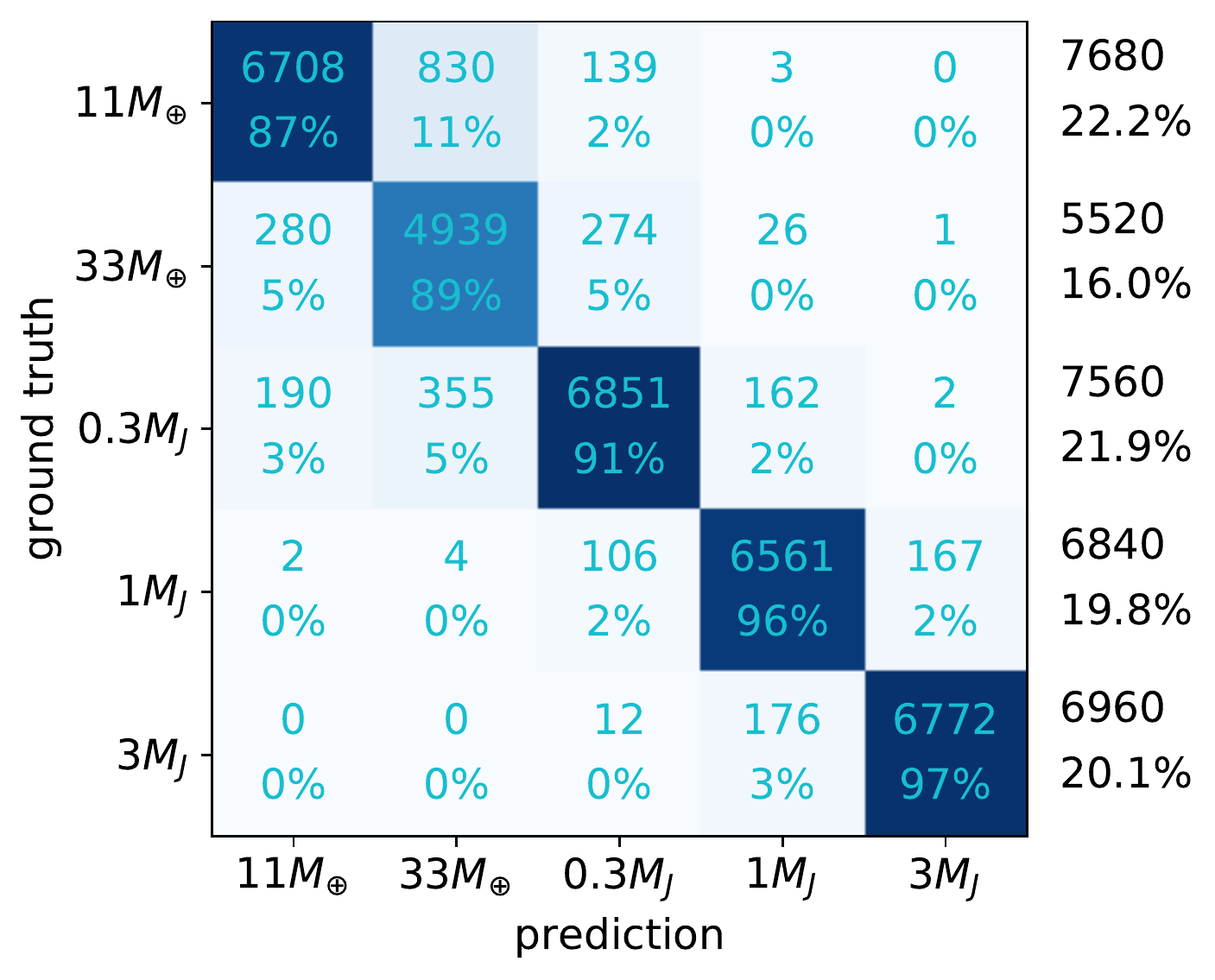}\caption{
\label{fig:confusion_mat} The confusion matrix for the ResNet model. The x-axis is the prediction from the neural network. The ground truth is on the y-axis. The upper number in a box shows the counts of planets with certain prediction and ground truth. The lower percentage shows the fraction of the prediction over the total number of the sample with certain ground truth (sum of a row). The rightmost numbers are total counts of testing data with certain class labels (sum of a given row) and the percentage among all testing data (sum of the rightmost column).}
\end{figure}

Figure \ref{fig:confusion_mat} shows the confusion matrix of planet mass prediction, and their ground truth for all testing data set applied to the ResNet network. The y-axis shows the true planet masses, whereas the x-axis shows the most likely planet masses predicted by the network. If the data point falls on the matrix diagonal, the prediction is correct. The accuracy is the sum of diagonal counts over the total counts. The planet mass is underestimated if it is on the left and overestimated if on the right. The upper number in each box shows the counts of a certain prediction given a ground truth. The lower percentage shows the fraction of it among all the images within that class.  When the planet mass is small, i.e., the gap is narrow, the prediction accuracy is low. If the planet is 11 $M_\oplus$, only 87\% of the samples account for correct predictions within that class. The planet mass is always overestimated given this setup. However, we caution that the planet mass can also be underestimated in reality, since this neural network cannot find planet with mass lower than this limit. The prediction accuracy increases with higher planet masses. If the planet mass is 3 $M_J$, 97\% of the predictions within that class are correct. Likewise, 3 $M_J$ is the upper mass limit in this network, so any real planet mass with higher value will be underestimated.

\subsection{Regression}

\begin{figure*}
\includegraphics[width=\linewidth]{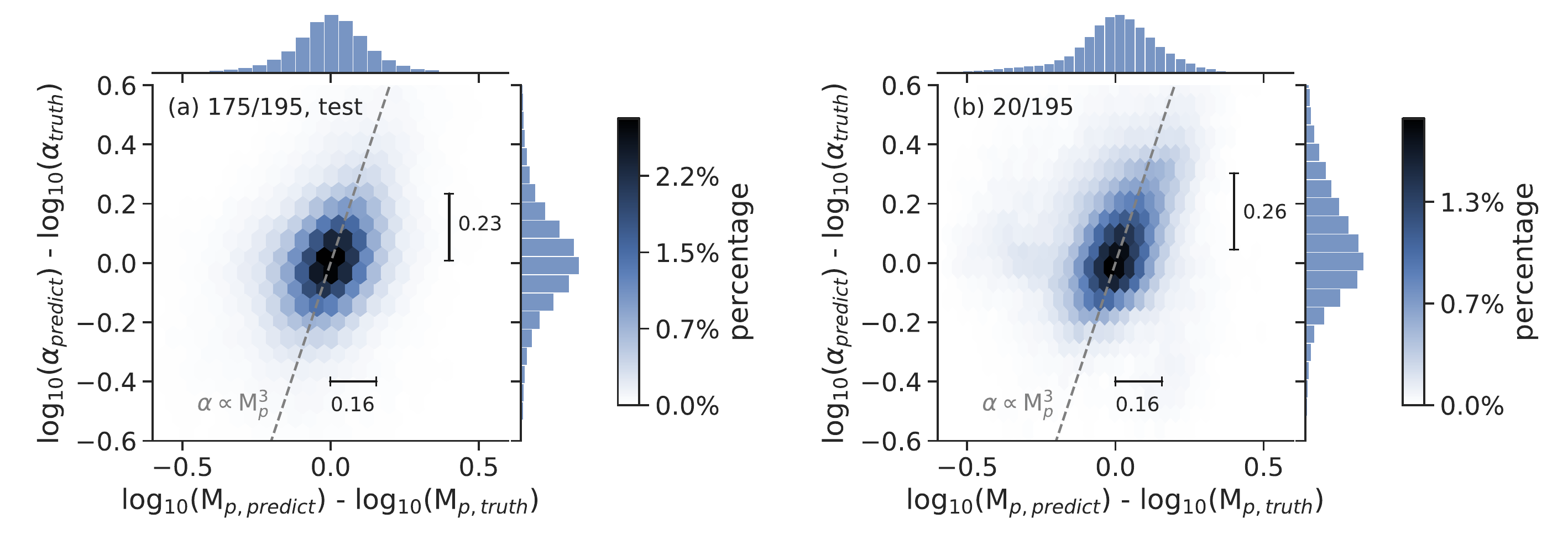}\caption{
\label{fig:jointplots}1D and 2D histograms for the differences between predicted and true values of $M_p$ (horizontal) and $\alpha$ (vertical). The 2D joint distributions are color-coded. (a) The test data set among 175/195 of the generic models. (b) The 20/195 data set that has not used in the training process. Most of the predictions have small deviations. The horizontal and vertical bars represent the standard deviation of each distributions. Dashed lines represent $\alpha \propto M_p^3$.
}
\end{figure*}

\begin{figure}
\includegraphics[width=\linewidth]{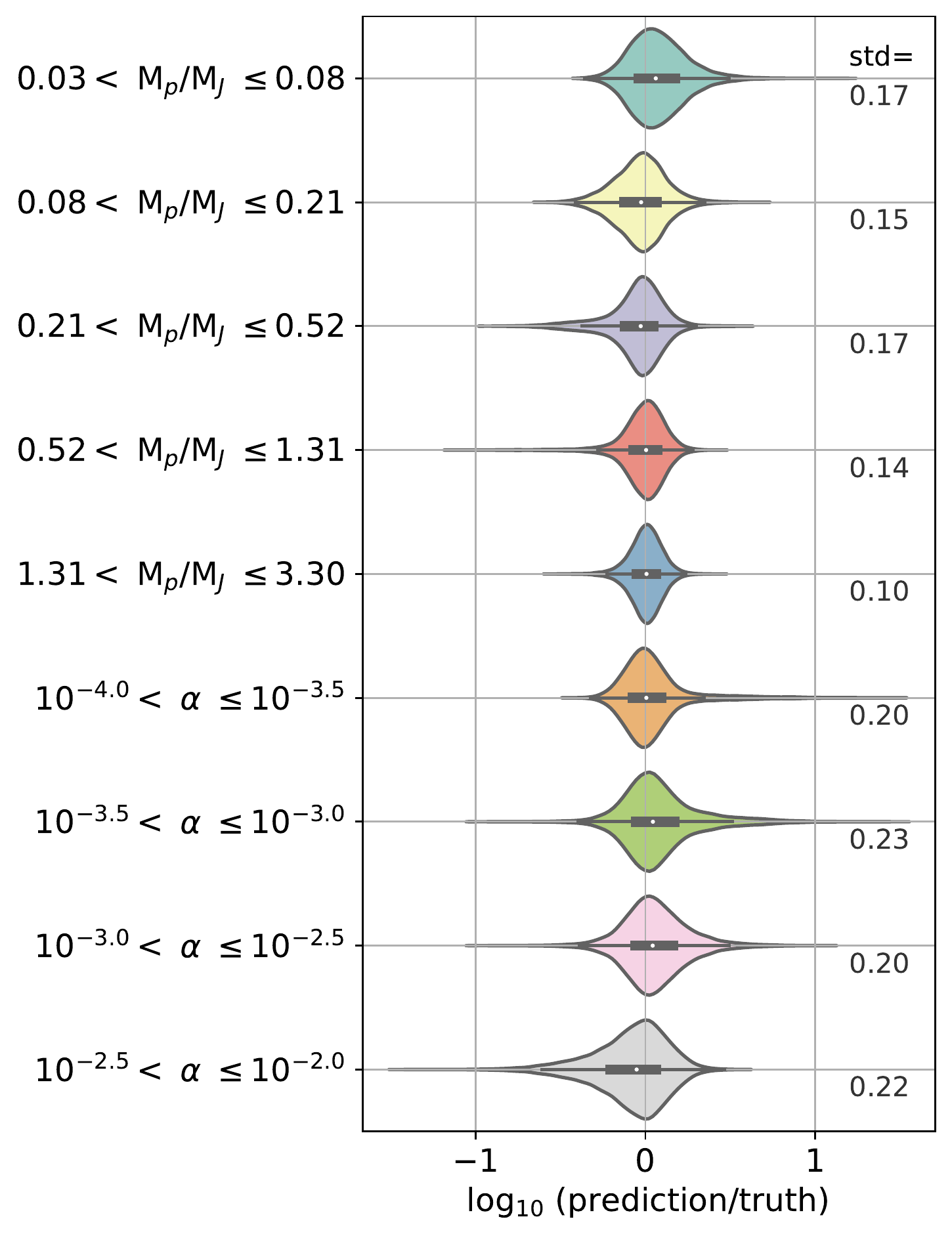}\caption{
\label{fig:violinplots}The violin plots of logarithmic-scaled deviation between predicted and true values of $M_p$ and $\alpha$ in different mass and viscosity regimes. The data are from the test set of 175/195 generic models. The color-shaded regions show the distribution with normalized height, whereas the inner box follows the convention of a box-plot, which shows distribution's 25 percentile, 50 percentile (white dot), and 75 percentile. The decimals marked on the right are the standard deviation of each distribution.
}
\end{figure}

The loss of the regression model reached a floor after 70 training epochs. Figure \ref{fig:jointplots} shows histograms for the deviation of the predicted planet masses and disk viscosities from the true values in dex. 2D histograms of the joint distribution between $M_p$ and $\alpha$ are plotted at center, whereas the 1D distributions of the planet mass and viscosity differences are shown on the top and right. Panel (a) shows the result of the test set among the 175/195 generic models that were used in the training. The distributions of $M_p$ and $\alpha$ are all centered close to zero, with uncertainties of 0.16 and 0.23 dex, respectively. They are symmetric with almost zero means. Planet masses and the disk viscosities tend to be over(under)-estimated at the same time and follow $\alpha \propto M_p^3$, which has also been found in \citet{zhang18}. Panel (b) shows the result of 20/195 generic models that are not used in the training. The overall distributions are similar to those in panel (a). The exception is that the deviation of $\alpha$ is skewed towards positive values (i.e., $\alpha$ tends to be over-predicted). This is because most of the viscosities among these randomly generated 20 generic models have low $\alpha$ (close to $10^{-4}$), where it is more likely to be overestimated (see Figure \ref{fig:violinplots}). The similarity between panel (a) and (b) demonstrates that the fitting is robust and can also be applied to simulations not used in the training.

Figure \ref{fig:violinplots} shows distributions of the difference of the prediction and ground truth in different mass and viscosity regimes. The data are from the test set of 175/195 generic models. The distributions are normalized to have the same height, and the inner box follows the convention of the box plot. The standard deviation of each distribution is listed on the right of each violin plot. The planet masses are divided into five bins in comparison with the classification problem (Figure \ref{fig:confusion_mat}). Similar to the classification problem, the error of the fitting becomes smaller as the planet mass increases. A small fraction of samples have large errors except at the highest mass bin. At the lower mass end, the planet mass is more likely to be overestimated. At the higher mass end, the uncertainty of the estimate can be as low as 0.1 dex (a factor of 1.3). The viscosity is divided into four bins. When $\alpha$ is less than 3$\times$ $10^{-3}$, it tends to be overestimated. When it is large and close to $10^{-2}$, it is more likely to be underestimated. In any viscosity regimes, a small fraction of the predictions would have large deviations from their true values.

\subsection{Grad-CAM}
While deep neural networks are difficult to interpret, some visualization tools help make sense of them. Gradient-weighted Class Activation Mapping (Grad-CAM; \citealt{selvaraju17}) is a technique for making CNN-based models more transparent by visualizing the regions of input that are important for predictions from these models. It uses the class-specific gradient information flowing into the final convolutional layer of a CNN to provide a coarse localization map of the important regions in the image. It is a generalization of the Class Activation Mapping \citep{zhou16}, but requires no re-training.

\begin{figure*}
\includegraphics[width=\linewidth]{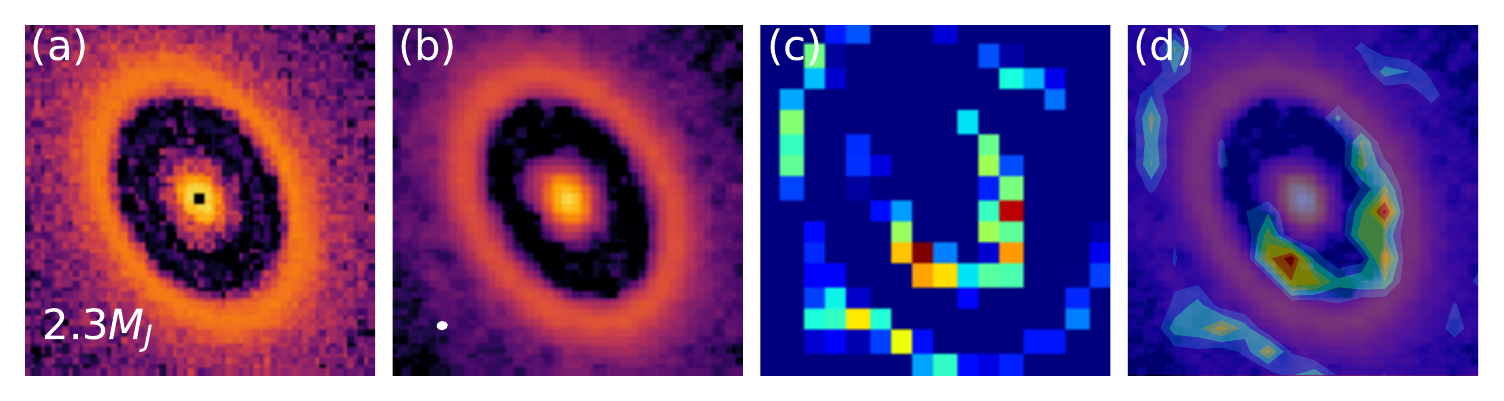}\caption{Grad-CAM for the ResNet regression problem. The image is among 20 generic models not in the training set. The true planet mass is 2.28 $M_J$ and $\alpha$ = 1.4 $\times$ 10$^{-3}$. (a) log-scaled image without putting into \texttt{CASA}, (b) log-scaled synthetic image using \texttt{CASA}, (c) activation map of the synthetic image (b), (d) synthetic image overlaid by the activation map (filled contour). In this case, the activation map successfully focuses on the gap region. The beam size of panel (b) is shown as an ellipse in the bottom left.
\label{fig:gradcam}}
\end{figure*}

Figure \ref{fig:gradcam} shows both images and the activation map derived from the Grad-CAM of the ResNet regression network. The disk is inclined and rotated. Here the true planet mass is 2.28 $M_J$, and $\alpha$ = 1.4 $\times$ 10$^{-3}$. The gap region has high values in the activation map, which means that it is important in predicting the planet mass. This lends confidence to the network's reliability since the traditional method also focuses on the gap's properties. This is why we named these neural networks as Planet Gap neural networks (PGNets). 
One should be cautious that not every activation map of testing data shows such a good correspondence, and the activation map is a great tool if any result is in doubt. 

\subsection{Application to observational data}

After testing against the synthetic observations, we applied our trained PGNets to real DSHARP observations (AS 209 \citealt{andrews18b, guzman18}; Elias 24, GW Lup and Sz 114; \citealt{andrews18b}) to infer planet masses and compared them with what found in \citet{zhang18}. The designation follows \citet{huang18b}, where the integer of gap location in au follows ``D'' (for dark gaps). There are several degeneracies if one wants to infer the planet mass. In the previous fitting of \cite{zhang18}, the gap width depends on not only the planet mass but also the viscosity $\alpha$ and Stokes number $St$ (or $a_{max}$ and $\Sigma_{g}$). Thus, in comparison with \cite{zhang18}, we picked the median value as a reference point. The median value is the case with $\alpha$=$10^{-3}$ and $a_{max}$=1 mm. It is difficult to estimate the uncertainty in these models. For the linear fitting method, two known uncertainties come from $\alpha$ and $a_{max}$. A difference of 10 in $\alpha$ leads to around 0.33 dex change of $M_p$, whereas a change of 10 in $a_{max}$ leads to around 0.2 dex change of $M_p$. For classification models, an error of reference can be 0.4 dex, since two neighbouring planet masses are spaced at that value. For the regression model, the uncertainty of $M_p$ and $\alpha$ can be estimated as 0.16 dex and 0.23 dex, as shown in Figure \ref{fig:jointplots}, or read in specific regimes as shown in Figure \ref{fig:violinplots}. For AS 209 and Elias 24, fine-tuned planet masses are available from detailed case-by-case modelings. While the central stars of these disks have different masses, what matters is the planet-star mass ratio $M_p/M_*$. Thus, we will report $M_p/M_*$ in units of the $M_J/M_\odot$ in the next paragraph. Notice that the radius of the semi-minor axis of the gap should be placed around one-sixth of the image size. However, the disk does not need to be centered nor scaled perfectly, since the images were randomly shifted during the training process.

AS 209 is a disk with many gaps. With the linear fitting method, the median value of AS 209 D99\footnote{It is actually B99 (``B'' stands for bright rings). However, that bright ring is just a shallow peak inside a wide gap.} is 0.45 $M_J/M_\odot$ \citep{zhang18}. The detailed modeling infers 0.1 $M_J/M_\odot$. The VGG-like model predicts 0.1 $M_J/M_\odot$, whereas the ResNet classification model predicts 0.3 $M_J/M_\odot$. The regression model predicts 0.52  $M_J/M_\odot$ and $\alpha = 2\times 10^{-4}$.  With 64$\times$64 resolution, gaps inside 40 au have been smoothed out. The secondary gap (a shallower gap inside the major gap) has already been considered since there are also many low-viscosity simulations with secondary gaps in our training data set.

With the traditional fitting method, Elias 24 D57 is 0.51 $M_J/M_\odot$. It is 0.2 $M_J/M_\odot$ using the fine-tuned model. Both VGG-like and ResNet classification model predicts 0.1 $M_J/M_\odot$. The regression model predicts $M_p/M_*$ = 0.21 $M_J/M_\odot$, and $\alpha = 2\times 10^{-4}$.

The inferred planet mass from the traditional fitting method for GW Lup D74 is 0.065 $M_J/M_\odot$. Both VGG-like and ResNet classification model predicts 0.03 $M_J/M_\odot$. The regression model predicts $M_p/M_*$ = 0.05 $M_J/M_{\odot}$, and $\alpha = 1\times 10^{-4}$. 

Sz 114 is a disk with a very narrow gap. It is too narrow to use the linear fitting method. The mass was obtained by directly comparing with simulations \citep{zhang18}. That value for Sz 114 D39 is 0.12 $M_J/M_\odot$. Both VGG-like and ResNet models predict 0.03 $M_J/M_\odot$. The regression model predicts $M_p/M_*$ = 0.11 $M_J/M_\odot$, and $\alpha = 6\times 10^{-4}$. 

Overall, the VGG-like and ResNet classification models predict similar planet masses. The regression model predicts higher planet masses. It predicts lower viscosities for these disks.

There are many other gaps that can be compared between the linear fitting method and PGNets. Instead of doing extensive coverage of all available gaps, we made our code available online so that users can apply these methods to any new or archival data. With a super-resolution technique, \citet{jennings21} even found many more gaps inside 30 au for DSHARP observations. Those images and radial profiles can also be inputs for our models using either the PGNets or the linear fitting tools.

\section{Discussion}\label{sec:discussion}

\subsection{Synthetic observation \label{sec:CASA}}
To test if predictions from CNN models are affected by the way we produce synthetic observations, we picked the case in Figure \ref{fig:gradcam} (a), put it into \texttt{CASA} \citep{mcmullin07} version 6.1.0, and used \texttt{simobserve} task to generate observations with angular resolutions and sensitivities comparable to those of the DSHARP observations.
We assumed the planet is at 40 au, $a_{max}$ = 1cm, $\Sigma_{g,0}$ = 30 $\mathrm{g/cm^2}$, and $L_*$ = 2 $L_\odot$. The synthetic observations consist of 12 minutes of on-source integration time with the Cycle 5 C43-5 antenna configuration, 35 minutes on-source in the C43-8 configuration, and 35 minutes on-source in the C43-9 configuration. A precipitable water vapor level of 1.0 mm was adopted. The synthetic visibility was imaged in the same manner as the DSHARP sources using \texttt{tclean} task, as described in \cite{andrews18b}.

In Figure \ref{fig:gradcam}, panel (a) shows the intensity map converted from the simulation and panel (b) shows the intensity from the synthetic ALMA observation. The scales of them were normalized as discussed in Section \ref{sec:preprocessing}. The angular resolution is $\sim$ 5 au in FWHM and is marked in the lower left corner of the panel. As shown in panel (c) and (d), the activated regions correctly focus on the gap. The predicted masses are 1.92 $M_J$ and 2.39 $M_J$, with and without $\texttt{CASA}$ operations, all comparable to the true value, 2.28 $M_J$. The predicted $\alpha$ viscosities are 1.3$\times 10^{-3}$ and 1.2$\times 10^{-3}$, with and without $\texttt{CASA}$ operations, also comparable to the true value, 1.4 $\times 10^{-3}$. Thus, we conclude that our models can be used to predict real ALMA observations even though they were trained on the data without doing \texttt{simobserve} tasks.

\subsection{Advantages}
Compared to the traditional method, CNNs have many advantages.
It is quick and convenient. The 2D images do not need to be converted to 1D profiles. Deriving 1D profiles seems to be a simple task but it takes time since one needs to run MCMC to find a disk's centroid, inclination, and position angle. Once the CNN model is trained, one can obtain a prediction within a second directly from the image plane. For this reason, it can be applied to a large disk sample. The training is one time and only takes an hour. Using a traditional method, one needs to search for the best fitting parameters. For instance, one parameter combining viscosity, $h/r$, and planet mass is enough to fit the gaseous gap depth and width \citep{kanagawa_depth, kanagawa_width}. However, if the dust is included, we also need to find several fitting formulae for different $a_{max}$ and gas surface density \citep{zhang18}. This is not the case for CNNs. With more data or new physics included, the network can be retrained quickly by providing more training data.

Asymmetric information is lost when a 2D image is converted to a 1D radial profile. CNNs can preserve this information. For instance, a low mass planet in an inviscid disk can have the same gap width as a high mass planet in a viscous disk. The traditional method cannot break this degeneracy. However, the viscosity of the disk can be inferred from a 2D image. If the disk looks more asymmetric, it should be more inviscid. The planet mass can be better constrained accordingly.

In a regression problem, the model can provide planet mass and disk viscosity at the same time. With the linear fitting method, one can only get a planet mass by assuming a disk viscosity, since both a higher planet mass and a more inviscid disk help open a wider gap. The CNN regression model partially breaks the degeneracy between them.

The CNN methods can handle very shallow or narrow gaps, which is of great difficulty for the traditional fitting method. In \citet{zhang18}, to make sure most of the data points fall onto the fitting line, narrow gaps with $\Delta$ below 0.15 were treated as outliers. One thus needs to compare with individual simulations to estimate a planet mass (e.g., inferring the potential planet in Sz 114's narrow gap), which is extremely time consuming.

\subsection{Limitations}
The first limitation of the CNN models is that training with the same data set but different architectures would lead to different predictions, even though their accuracy is similar statistically.  Even given the same architecture, different random seeds or data augmentation would also result in different predictions. On the contrary, the traditional linear fitting method provides a definite planet mass as long as a gap width and other disk parameters are provided. Many details of fitting are empirical, but they are transparent to the user.

Compared to the linear fitting method, it is difficult to understand how CNN methods derive the results, even though the Grad-CAM can qualitatively inform us the important features the network uses. In the traditional fitting method, the gap width is proportional to some powers of planet mass, disk viscosity, and $h/r$ at a given $St$. We can use some other ways to constrain some parameters and then narrow down the planet mass. In classification or single-output regression CNN models, we can only get a single prediction for these parameters when we apply them to observations. Deep neural networks are intrinsically highly non-linear. For an input image, these models are more of methods that help pick up an image in a training set with the most similar feature than methods of finding intrinsic relationships. The user can use their own knowledge to interpret why the planet mass returned by a CNN model is as such (e.g., if the planet mass is low, the user can make sense of it by noticing the gap is narrow.), but the network itself cannot inform this to the user. Instead of using a linear fitting with many assumptions, it helps the user to find the most closed-match model for an input. In some sense, the classification and single-output regression models can be seen as another way to present all the simulations in \cite{zhang18} by providing an automated tool to find a closed-match. Surprisingly, multi-output regression models can possibly solve this problem. For instance, the multi-output regression PGNets can help us find the exact relation of the degeneracy between planet mass and disk viscosity among simulations (Figure \ref{fig:jointplots}). There are hopes that we can learn valuable insights from CNN models.

Finally, compared to a detailed modelling, the prediction can only be as good as the physics included in the simulations. While our simulations span a large parameter space, they cannot cover every possible situation. For instance, we cannot predict planet masses below 11 $M_\oplus$ or above 3 $M_J$. A planet with a lower mass than 11 $M_\oplus$ can only be predicted as massive as that. Only one planet is put into the simulation, and its orbit is fixed. Thus, the model cannot be used to study multiple planets carving a common gap. However, if planets lead to several gaps and they do not influence each other, we can treat them as individual single gaps by masking others. Note that if we assume that one planet can carve two gaps \citep{dong18b,bae2018a, bae2018b} and the secondary gap is at 0.5 - 0.7 r$_p$, we can input all the gaps into CNN models since our CNN models have been trained with data which have these secondary gaps generated by a single planet. The model does not consider migration, which can lead to a different gap shape \citep{Nazari2019,kanagawa20}.
The gap substructure can also change with time. The disks are at 1000 orbits, which is 1 Myr for a planet at 100 au or 0.1 Myr for a planet at 20 au. 
We used 2D simulations, but the situation in 3D might be different. We neglected the self-gravity of the disk and the thermodynamic processes. These effects will also change the shape of the gap \citep{ zhang20, miranda20a, miranda20b, ziampras20, Rowther2020}. The dust in our simulations was treated as passive test particles. In reality, the dust's back-reaction onto the gas is important \citep{Kanagawa2018,yang20,Huang2020,Hsieh2020}. 
When producing the synthetic image, we neglected scattering, but it can affect the images when the disk is optically thick \citep{Liu2019,zhu19b}. The dust was also assumed to be settled. 
Not to mention that it is very likely that some substructures can have non-planet origins (e.g., snowline and MHD effects). 
On the other hand, with more physical processes understood and included in simulations, CNNs can be used to make predictions. 

\subsection{Future perspectives}
This line of work can also be applied to infer the Stokes number $St$ of particles in the disks. The Stokes number is highly correlated to the gap width and depth.  Measuring $St$ can help us understand the particle size and dust settling.

The gas component is more massive and has a larger radial and vertical extent than the dust in protoplanetary disks. The (sub)mm molecular line observations also contain velocity information at different disk positions, leading to a 3D datacube. One can infer planet mass from kinematic features in channel maps, such as the kink and the deviation from Keplerian velocity \citep{pinte2018, teague18a}. In star formation, CNN models have already been used on the whole 3D datacube \citep{xu20a, xu20b}. This can possibly be applicable in protoplanetary disks as well.

\section{Conclusions}\label{sec:conclusion}
Substructures are found to be ubiquitous in protoplanetary disks. If some of them are induced by planets, the increasing number of high-resolution protoplanetary disk observations is revealing the population of young forming planets. 

The properties of these substructures (e.g., the width and depth of the gap) are related to the planet mass. Previous works used either fined-tuned models or linear fitting on a large parameter space to infer the planet mass. Instead, we used Convolutional Neural Networks to predict the planet mass directly from radio dust continuum images. To train the CNNs, we use data from synthetic observations in \citet{zhang18} and some new simulations. We built both classification and regression models. The classification models can predict five planet masses ranging from 11 $M_\oplus$ to 3 $M_J$. The VGG-like model can reach 89\% accuracy, whereas the ResNet model can reach 92\% accuracy. The accuracies for less massive planets are lower. They are higher for more massive planets. The regression model can predict planet mass and disk viscosity at the same time. Similar to the classification model, it predicts more massive planets with higher accuracy. The standard deviation of the prediction is around 0.16 dex. The prediction of $\alpha$ has uncertainty around 0.23 dex, and can be used to constrain the disk turbulence. 

CNN models cannot fully break the degeneracy between the planet mass and disk viscosity. It tends to over(under)-predict them at the same time. However, it is surprising that this degeneracy relationship can be easily found in CNNs without the need of theoretical knowledge \citep{kanagawa_width} and detailed fitting and tuning \citep{zhang18}. This shows the potential of CNNs as diagnostic tools.

Using Grad-CAM, we showed that the networks indeed catch the important feature, i.e., gaps, in predicting the planet mass. We also applied the networks to several DSHARP gaps and found the predictions are reasonable compared with the traditional method in \citet{zhang18}. The code, along with that of the traditional method, is also provided.

The CNN methods are fast compared to fined-tuned models. It is more convenient than the linear fitting method since one can get a prediction instantaneously as long as an image is provided. The network can also be easily updated with more data or physical processes. It preserves the 2D information that should help break the degeneracy between planet mass and disk properties. The regression model can predict several quantities at the same time. Unlike the traditional method, predicting shallow and narrow gaps takes the same amount of effort using CNNs, even though the prediction still has higher uncertainty in this regime.

There are also several shortcomings for CNN models. Different architectures or training procedures might lead to different predictions. The CNNs are not transparent, and it is difficult to know how exactly the networks work. Lastly, in contrast to the detailed modelling, the robustness of our CNNs ultimately is limited by our training data (simulations), i.e., the physical processes included. 

The methods are more suitable for a large disk sample to obtain statistical trends of a young planet population. For individual disk, we can use this method to narrow down the parameter space for detailed simulations. Overall, the traditional linear fitting method (e.g., \citealt{zhang18}) still provides users more control on the input disk parameters, while the wide choice of tools in CNNs (e.g., classification and regression) and diagnostic tools (Grad-CAM) start to make CNNs more robust.

\section*{Acknowledgements}
We thank the anonymous reviewer who raised insightful questions that significantly improved the quality of this manuscript. SZ thanks Duo Xu for helpful discussions. SZ acknowledges support from the University of Nevada, Las Vegas Barrick Fellowship. ZZ acknowledges support from the National Science Foundation under CAREER Grant No. AST-1753168.

\section*{Data availability}
The data underlying this article are public available on GitHub at \url{https://github.com/zhangsj96/DSHARPVII.git} for the linear fitting method, and \url{https://github.com/zhangsj96/PGNets.git} for convolutional neural networks.

\section*{Software}
{\tt Dusty FARGO-ADSG} \citep{baruteau2008a, baruteau2008b, baruteau2016},
{\tt CASA} \citep{mcmullin07},
{\tt Tensorflow} \citep{tensorflow2015-whitepaper},
{\tt Matplotlib} \citep{matplotlib},
{\tt Numpy} \citep{numpy}, 
{\tt Scipy} \citep{scipy},
{\tt Seaborn} \citep{seaborn},
{\tt Astropy} \citep{astropy}




\bibliographystyle{mnras}

\begin{thebibliography}{}
\makeatletter
\relax
\def\mn@urlcharsother{\let\do\@makeother \do\$\do\&\do\#\do\^\do\_\do\%\do\~}
\def\mn@doi{\begingroup\mn@urlcharsother \@ifnextchar [ {\mn@doi@}
  {\mn@doi@[]}}
\def\mn@doi@[#1]#2{\def\@tempa{#1}\ifx\@tempa\@empty \href
  {http://dx.doi.org/#2} {doi:#2}\else \href {http://dx.doi.org/#2} {#1}\fi
  \endgroup}
\def\mn@eprint#1#2{\mn@eprint@#1:#2::\@nil}
\def\mn@eprint@arXiv#1{\href {http://arxiv.org/abs/#1} {{\tt arXiv:#1}}}
\def\mn@eprint@dblp#1{\href {http://dblp.uni-trier.de/rec/bibtex/#1.xml}
  {dblp:#1}}
\def\mn@eprint@#1:#2:#3:#4\@nil{\def\@tempa {#1}\def\@tempb {#2}\def\@tempc
  {#3}\ifx \@tempc \@empty \let \@tempc \@tempb \let \@tempb \@tempa \fi \ifx
  \@tempb \@empty \def\@tempb {arXiv}\fi \@ifundefined
  {mn@eprint@\@tempb}{\@tempb:\@tempc}{\expandafter \expandafter \csname
  mn@eprint@\@tempb\endcsname \expandafter{\@tempc}}}

\bibitem[\protect\citeauthoryear{Abadi et~al.,}{Abadi
  et~al.}{2015}]{tensorflow2015-whitepaper}
Abadi M.,  et~al., 2015, {TensorFlow}: Large-Scale Machine Learning on
  Heterogeneous Systems, \url {https://www.tensorflow.org/}

\bibitem[\protect\citeauthoryear{{Alibert} \& {Venturini}}{{Alibert} \&
  {Venturini}}{2019}]{alibert19}
{Alibert} Y.,  {Venturini} J.,  2019, \mn@doi [\aap]
  {10.1051/0004-6361/201834942}, \href
  {https://ui.adsabs.harvard.edu/abs/2019A&A...626A..21A} {626, A21}

\bibitem[\protect\citeauthoryear{{Andrews} et~al.,}{{Andrews}
  et~al.}{2018}]{andrews18b}
{Andrews} S.~M.,  et~al., 2018, \mn@doi [\apjl] {10.3847/2041-8213/aaf741},
  \href {http://adsabs.harvard.edu/abs/2018ApJ...869L..41A} {869, L41}

\bibitem[\protect\citeauthoryear{{Astropy Collaboration} et~al.,}{{Astropy
  Collaboration} et~al.}{2013}]{astropy}
{Astropy Collaboration} et~al., 2013, \mn@doi [\aap]
  {10.1051/0004-6361/201322068}, \href
  {http://adsabs.harvard.edu/abs/2013A%26A...558A..33A} {558, A33}

\bibitem[\protect\citeauthoryear{{Auddy} \& {Lin}}{{Auddy} \&
  {Lin}}{2020}]{auddy20}
{Auddy} S.,  {Lin} M.-K.,  2020, \mn@doi [\apj] {10.3847/1538-4357/aba95d},
  \href {https://ui.adsabs.harvard.edu/abs/2020ApJ...900...62A} {900, 62}

\bibitem[\protect\citeauthoryear{{Auddy}, {Dey}, {Lin}  \& {Hall}}{{Auddy}
  et~al.}{2021}]{Auddy2021}
{Auddy} S.,  {Dey} R.,  {Lin} M.-K.,   {Hall} C.,  2021, \mn@doi [\apj]
  {10.3847/1538-4357/ac1518}, \href
  {https://ui.adsabs.harvard.edu/abs/2021ApJ...920....3A} {920, 3}

\bibitem[\protect\citeauthoryear{{Bae} \& {Zhu}}{{Bae} \&
  {Zhu}}{2018a}]{bae2018a}
{Bae} J.,  {Zhu} Z.,  2018a, \mn@doi [\apj] {10.3847/1538-4357/aabf8c}, \href
  {http://adsabs.harvard.edu/abs/2018ApJ...859..118B} {859, 118}

\bibitem[\protect\citeauthoryear{{Bae} \& {Zhu}}{{Bae} \&
  {Zhu}}{2018b}]{bae2018b}
{Bae} J.,  {Zhu} Z.,  2018b, \mn@doi [\apj] {10.3847/1538-4357/aabf93}, \href
  {http://adsabs.harvard.edu/abs/2018ApJ...859..119B} {859, 119}

\bibitem[\protect\citeauthoryear{{Baruteau} \& {Masset}}{{Baruteau} \&
  {Masset}}{2008a}]{baruteau2008a}
{Baruteau} C.,  {Masset} F.,  2008a, \mn@doi [\apj] {10.1086/523667}, \href
  {http://adsabs.harvard.edu/abs/2008ApJ...672.1054B} {672, 1054}

\bibitem[\protect\citeauthoryear{{Baruteau} \& {Masset}}{{Baruteau} \&
  {Masset}}{2008b}]{baruteau2008b}
{Baruteau} C.,  {Masset} F.,  2008b, \mn@doi [\apj] {10.1086/529487}, \href
  {http://adsabs.harvard.edu/abs/2008ApJ...678..483B} {678, 483}

\bibitem[\protect\citeauthoryear{{Baruteau} \& {Zhu}}{{Baruteau} \&
  {Zhu}}{2016}]{baruteau2016}
{Baruteau} C.,  {Zhu} Z.,  2016, \mn@doi [\mnras] {10.1093/mnras/stv2527},
  \href {http://adsabs.harvard.edu/abs/2016MNRAS.458.3927B} {458, 3927}

\bibitem[\protect\citeauthoryear{{Birnstiel} et~al.,}{{Birnstiel}
  et~al.}{2018}]{birnstiel18}
{Birnstiel} T.,  et~al., 2018, \mn@doi [\apjl] {10.3847/2041-8213/aaf743},
  \href {https://ui.adsabs.harvard.edu/abs/2018ApJ...869L..45B} {869, L45}

\bibitem[\protect\citeauthoryear{{Christiaens} et~al.,}{{Christiaens}
  et~al.}{2019}]{Christiaens2019}
{Christiaens} V.,  et~al., 2019, \mn@doi [\mnras] {10.1093/mnras/stz1232},
  \href {https://ui.adsabs.harvard.edu/abs/2019MNRAS.486.5819C} {486, 5819}

\bibitem[\protect\citeauthoryear{{Cieza} et~al.,}{{Cieza}
  et~al.}{2021}]{Cieza2021}
{Cieza} L.~A.,  et~al., 2021, \mn@doi [\mnras] {10.1093/mnras/staa3787}, \href
  {https://ui.adsabs.harvard.edu/abs/2021MNRAS.501.2934C} {501, 2934}

\bibitem[\protect\citeauthoryear{{Dieleman}, {Willett}  \& {Dambre}}{{Dieleman}
  et~al.}{2015}]{dieleman15}
{Dieleman} S.,  {Willett} K.~W.,   {Dambre} J.,  2015, \mn@doi [\mnras]
  {10.1093/mnras/stv632}, \href
  {https://ui.adsabs.harvard.edu/abs/2015MNRAS.450.1441D} {450, 1441}

\bibitem[\protect\citeauthoryear{{Dipierro} \& {Laibe}}{{Dipierro} \&
  {Laibe}}{2017}]{dipierro2017}
{Dipierro} G.,  {Laibe} G.,  2017, \mn@doi [\mnras] {10.1093/mnras/stx977},
  \href {https://ui.adsabs.harvard.edu/abs/2017MNRAS.469.1932D} {469, 1932}

\bibitem[\protect\citeauthoryear{{Dipierro}, {Price}, {Laibe}, {Hirsh},
  {Cerioli}  \& {Lodato}}{{Dipierro} et~al.}{2015}]{Dipierro2015a}
{Dipierro} G.,  {Price} D.,  {Laibe} G.,  {Hirsh} K.,  {Cerioli} A.,   {Lodato}
  G.,  2015, \mn@doi [\mnras] {10.1093/mnrasl/slv105}, \href
  {https://ui.adsabs.harvard.edu/abs/2015MNRAS.453L..73D} {453, L73}

\bibitem[\protect\citeauthoryear{{Dipierro} et~al.,}{{Dipierro}
  et~al.}{2018}]{dipierro2018}
{Dipierro} G.,  et~al., 2018, \mn@doi [\mnras] {10.1093/mnras/sty181}, \href
  {http://adsabs.harvard.edu/abs/2018MNRAS.475.5296D} {475, 5296}

\bibitem[\protect\citeauthoryear{{Dong}, {Zhu}  \& {Whitney}}{{Dong}
  et~al.}{2015}]{dong2015a}
{Dong} R.,  {Zhu} Z.,   {Whitney} B.,  2015, \mn@doi [\apj]
  {10.1088/0004-637X/809/1/93}, \href
  {https://ui.adsabs.harvard.edu/abs/2015ApJ...809...93D} {809, 93}

\bibitem[\protect\citeauthoryear{{Dong}, {Li}, {Chiang}  \& {Li}}{{Dong}
  et~al.}{2018}]{dong18b}
{Dong} R.,  {Li} S.,  {Chiang} E.,   {Li} H.,  2018, \mn@doi [\apj]
  {10.3847/1538-4357/aadadd}, \href
  {https://ui.adsabs.harvard.edu/abs/2018ApJ...866..110D} {866, 110}

\bibitem[\protect\citeauthoryear{{Flaherty}, {Hughes}, {Teague}, {Simon},
  {Andrews}  \& {Wilner}}{{Flaherty} et~al.}{2018}]{flaherty2018}
{Flaherty} K.~M.,  {Hughes} A.~M.,  {Teague} R.,  {Simon} J.~B.,  {Andrews}
  S.~M.,   {Wilner} D.~J.,  2018, \mn@doi [\apj] {10.3847/1538-4357/aab615},
  \href {http://adsabs.harvard.edu/abs/2018ApJ...856..117F} {856, 117}

\bibitem[\protect\citeauthoryear{{Fung}, {Shi}  \& {Chiang}}{{Fung}
  et~al.}{2014}]{fung14}
{Fung} J.,  {Shi} J.-M.,   {Chiang} E.,  2014, \mn@doi [\apj]
  {10.1088/0004-637X/782/2/88}, \href
  {http://adsabs.harvard.edu/abs/2014ApJ...782...88F} {782, 88}

\bibitem[\protect\citeauthoryear{{Guzm{\'a}n} et~al.,}{{Guzm{\'a}n}
  et~al.}{2018}]{guzman18}
{Guzm{\'a}n} V.~V.,  et~al., 2018, \mn@doi [\apjl] {10.3847/2041-8213/aaedae},
  \href {https://ui.adsabs.harvard.edu/abs/2018ApJ...869L..48G} {869, L48}

\bibitem[\protect\citeauthoryear{{Haffert}, {Bohn}, {de Boer}, {Snellen},
  {Brinchmann}, {Girard}, {Keller}  \& {Bacon}}{{Haffert}
  et~al.}{2019}]{Haffert2019}
{Haffert} S.~Y.,  {Bohn} A.~J.,  {de Boer} J.,  {Snellen} I.~A.~G.,
  {Brinchmann} J.,  {Girard} J.~H.,  {Keller} C.~U.,   {Bacon} R.,  2019,
  \mn@doi [Nature Astronomy] {10.1038/s41550-019-0780-5}, \href
  {https://ui.adsabs.harvard.edu/abs/2019NatAs...3..749H} {3, 749}

\bibitem[\protect\citeauthoryear{{Hashimoto}, {Aoyama}, {Konishi}, {Uyama},
  {Takasao}, {Ikoma}  \& {Tanigawa}}{{Hashimoto} et~al.}{2020}]{Hashimoto2020}
{Hashimoto} J.,  {Aoyama} Y.,  {Konishi} M.,  {Uyama} T.,  {Takasao} S.,
  {Ikoma} M.,   {Tanigawa} T.,  2020, \mn@doi [\aj] {10.3847/1538-3881/ab811e},
  \href {https://ui.adsabs.harvard.edu/abs/2020AJ....159..222H} {159, 222}

\bibitem[\protect\citeauthoryear{{Hassan}, {Liu}, {Kohn}  \& {La
  Plante}}{{Hassan} et~al.}{2019}]{hassan19}
{Hassan} S.,  {Liu} A.,  {Kohn} S.,   {La Plante} P.,  2019, \mn@doi [\mnras]
  {10.1093/mnras/sty3282}, \href
  {https://ui.adsabs.harvard.edu/abs/2019MNRAS.483.2524H} {483, 2524}

\bibitem[\protect\citeauthoryear{He, Zhang, Ren  \& Sun}{He
  et~al.}{2016a}]{he16b}
He K.,  Zhang X.,  Ren S.,   Sun J.,  2016a, in Leibe B.,  Matas J.,  Sebe N.,
   Welling M.,  eds, Computer Vision -- ECCV 2016. Springer International
  Publishing, Cham, pp 630--645

\bibitem[\protect\citeauthoryear{He, Zhang, Ren  \& Sun}{He
  et~al.}{2016b}]{he16a}
He K.,  Zhang X.,  Ren S.,   Sun J.,  2016b, \mn@doi [Proceedings of the IEEE
  Computer Society Conference on Computer Vision and Pattern Recognition]
  {10.1109/CVPR.2016.90}, 2016-December, 770

\bibitem[\protect\citeauthoryear{{Hezaveh}, {Perreault Levasseur}  \&
  {Marshall}}{{Hezaveh} et~al.}{2017}]{hezaveh17}
{Hezaveh} Y.~D.,  {Perreault Levasseur} L.,   {Marshall} P.~J.,  2017, \mn@doi
  [\nat] {10.1038/nature23463}, \href
  {https://ui.adsabs.harvard.edu/abs/2017Natur.548..555H} {548, 555}

\bibitem[\protect\citeauthoryear{{Hsieh} \& {Lin}}{{Hsieh} \&
  {Lin}}{2020}]{Hsieh2020}
{Hsieh} H.-F.,  {Lin} M.-K.,  2020, \mn@doi [\mnras] {10.1093/mnras/staa2115},
  \href {https://ui.adsabs.harvard.edu/abs/2020MNRAS.497.2425H} {497, 2425}

\bibitem[\protect\citeauthoryear{{Huang} et~al.,}{{Huang}
  et~al.}{2018}]{huang18b}
{Huang} J.,  et~al., 2018, \mn@doi [\apjl] {10.3847/2041-8213/aaf740}, \href
  {https://ui.adsabs.harvard.edu/abs/2018ApJ...869L..42H} {869, L42}

\bibitem[\protect\citeauthoryear{{Huang}, {Li}, {Isella}, {Miranda}, {Li}  \&
  {Ji}}{{Huang} et~al.}{2020}]{Huang2020}
{Huang} P.,  {Li} H.,  {Isella} A.,  {Miranda} R.,  {Li} S.,   {Ji} J.,  2020,
  \mn@doi [\apj] {10.3847/1538-4357/ab8199}, \href
  {https://ui.adsabs.harvard.edu/abs/2020ApJ...893...89H} {893, 89}

\bibitem[\protect\citeauthoryear{Hunter}{Hunter}{2007}]{matplotlib}
Hunter J.~D.,  2007, Computing In Science \& Engineering, 9, 90

\bibitem[\protect\citeauthoryear{Ioffe \& Szegedy}{Ioffe \&
  Szegedy}{2015}]{ioffe15}
Ioffe S.,  Szegedy C.,  2015, in ICML. pp 448--456, \url
  {http://proceedings.mlr.press/v37/ioffe15.html}

\bibitem[\protect\citeauthoryear{{Isella}, {Benisty}, {Teague}, {Bae},
  {Keppler}, {Facchini}  \& {P{\'e}rez}}{{Isella} et~al.}{2019}]{Isella2019}
{Isella} A.,  {Benisty} M.,  {Teague} R.,  {Bae} J.,  {Keppler} M.,  {Facchini}
  S.,   {P{\'e}rez} L.,  2019, \mn@doi [\apjl] {10.3847/2041-8213/ab2a12},
  \href {https://ui.adsabs.harvard.edu/abs/2019ApJ...879L..25I} {879, L25}

\bibitem[\protect\citeauthoryear{{Izquierdo}, {Testi}, {Facchini}, {Rosotti}
  \& {van Dishoeck}}{{Izquierdo} et~al.}{2021}]{Izquierdo2021}
{Izquierdo} A.~F.,  {Testi} L.,  {Facchini} S.,  {Rosotti} G.~P.,   {van
  Dishoeck} E.~F.,  2021, \mn@doi [\aap] {10.1051/0004-6361/202140779}, \href
  {https://ui.adsabs.harvard.edu/abs/2021A&A...650A.179I} {650, A179}

\bibitem[\protect\citeauthoryear{{Jennings}, {Booth}, {Tazzari}, {Clarke}  \&
  {Rosotti}}{{Jennings} et~al.}{2021}]{jennings21}
{Jennings} J.,  {Booth} R.~A.,  {Tazzari} M.,  {Clarke} C.~J.,   {Rosotti}
  G.~P.,  2021, arXiv e-prints, \href
  {https://ui.adsabs.harvard.edu/abs/2021arXiv210302392J} {p. arXiv:2103.02392}

\bibitem[\protect\citeauthoryear{{Kanagawa}, {Muto}, {Tanaka}, {Tanigawa},
  {Takeuchi}, {Tsukagoshi}  \& {Momose}}{{Kanagawa}
  et~al.}{2015}]{kanagawa_depth}
{Kanagawa} K.~D.,  {Muto} T.,  {Tanaka} H.,  {Tanigawa} T.,  {Takeuchi} T.,
  {Tsukagoshi} T.,   {Momose} M.,  2015, \mn@doi [\apjl]
  {10.1088/2041-8205/806/1/L15}, \href
  {http://adsabs.harvard.edu/abs/2015ApJ...806L..15K} {806, L15}

\bibitem[\protect\citeauthoryear{{Kanagawa}, {Muto}, {Tanaka}, {Tanigawa},
  {Takeuchi}, {Tsukagoshi}  \& {Momose}}{{Kanagawa}
  et~al.}{2016}]{kanagawa_width}
{Kanagawa} K.~D.,  {Muto} T.,  {Tanaka} H.,  {Tanigawa} T.,  {Takeuchi} T.,
  {Tsukagoshi} T.,   {Momose} M.,  2016, \mn@doi [\pasj] {10.1093/pasj/psw037},
  \href {http://adsabs.harvard.edu/abs/2016PASJ...68...43K} {68, 43}

\bibitem[\protect\citeauthoryear{{Kanagawa}, {Muto}, {Okuzumi}, {Tanigawa},
  {Taki}  \& {Shibaike}}{{Kanagawa} et~al.}{2018}]{Kanagawa2018}
{Kanagawa} K.~D.,  {Muto} T.,  {Okuzumi} S.,  {Tanigawa} T.,  {Taki} T.,
  {Shibaike} Y.,  2018, \mn@doi [\apj] {10.3847/1538-4357/aae837}, \href
  {https://ui.adsabs.harvard.edu/abs/2018ApJ...868...48K} {868, 48}

\bibitem[\protect\citeauthoryear{{Kanagawa}, {Nomura}, {Tsukagoshi}, {Muto}  \&
  {Kawabe}}{{Kanagawa} et~al.}{2020}]{kanagawa20}
{Kanagawa} K.~D.,  {Nomura} H.,  {Tsukagoshi} T.,  {Muto} T.,   {Kawabe} R.,
  2020, \mn@doi [\apj] {10.3847/1538-4357/ab781e}, \href
  {https://ui.adsabs.harvard.edu/abs/2020ApJ...892...83K} {892, 83}

\bibitem[\protect\citeauthoryear{{Keppler} et~al.,}{{Keppler}
  et~al.}{2018}]{Keppler2018}
{Keppler} M.,  et~al., 2018, \mn@doi [\aap] {10.1051/0004-6361/201832957},
  \href {http://adsabs.harvard.edu/abs/2018A%26A...617A..44K} {617, A44}

\bibitem[\protect\citeauthoryear{Kingma \& Ba}{Kingma \& Ba}{2015}]{kingma14}
Kingma D.~P.,  Ba J.,  2015, in Bengio Y.,  LeCun Y.,  eds, 3rd International
  Conference on Learning Representations, {ICLR} 2015, San Diego, CA, USA, May
  7-9, 2015, Conference Track Proceedings. \url
  {http://arxiv.org/abs/1412.6980}

\bibitem[\protect\citeauthoryear{LeCun, Bottou, Bengio  \& Haffner}{LeCun
  et~al.}{1998}]{lecun98}
LeCun Y.,  Bottou L.,  Bengio Y.,   Haffner P.,  1998, \mn@doi [Proceedings of
  the IEEE] {10.1109/5.726791}, 86, 2278

\bibitem[\protect\citeauthoryear{{Lieu}, {Conversi}, {Altieri}  \&
  {Carry}}{{Lieu} et~al.}{2019}]{lieu19}
{Lieu} M.,  {Conversi} L.,  {Altieri} B.,   {Carry} B.,  2019, \mn@doi [\mnras]
  {10.1093/mnras/stz761}, \href
  {https://ui.adsabs.harvard.edu/abs/2019MNRAS.485.5831L} {485, 5831}

\bibitem[\protect\citeauthoryear{{Liu}}{{Liu}}{2019}]{Liu2019}
{Liu} H.~B.,  2019, \mn@doi [\apjl] {10.3847/2041-8213/ab1f8e}, \href
  {https://ui.adsabs.harvard.edu/abs/2019ApJ...877L..22L} {877, L22}

\bibitem[\protect\citeauthoryear{{Lodato} et~al.,}{{Lodato}
  et~al.}{2019}]{Lodato2019}
{Lodato} G.,  et~al., 2019, \mn@doi [\mnras] {10.1093/mnras/stz913}, \href
  {https://ui.adsabs.harvard.edu/abs/2019MNRAS.486..453L} {486, 453}

\bibitem[\protect\citeauthoryear{{Long} et~al.,}{{Long} et~al.}{2018}]{long18}
{Long} F.,  et~al., 2018, \mn@doi [The Astrophysical Journal]
  {10.3847/1538-4357/aae8e1}, \href
  {https://ui.adsabs.harvard.edu/abs/2018ApJ...869...17L} {869, 17}

\bibitem[\protect\citeauthoryear{McKay, Beckman  \& Conover}{McKay
  et~al.}{1979}]{McKay79}
McKay M.~D.,  Beckman R.~J.,   Conover W.~J.,  1979, Technometrics, 21, 239

\bibitem[\protect\citeauthoryear{{McMullin}, {Waters}, {Schiebel}, {Young}  \&
  {Golap}}{{McMullin} et~al.}{2007}]{mcmullin07}
{McMullin} J.~P.,  {Waters} B.,  {Schiebel} D.,  {Young} W.,   {Golap} K.,
  2007, in {Shaw} R.~A.,  {Hill} F.,   {Bell} D.~J.,  eds,  Astronomical
  Society of the Pacific Conference Series Vol. 376, Astronomical Data Analysis
  Software and Systems XVI. p.~127

\bibitem[\protect\citeauthoryear{{Miranda} \& {Rafikov}}{{Miranda} \&
  {Rafikov}}{2020a}]{miranda20a}
{Miranda} R.,  {Rafikov} R.~R.,  2020a, \mn@doi [\apj]
  {10.3847/1538-4357/ab791a}, \href
  {https://ui.adsabs.harvard.edu/abs/2020ApJ...892...65M} {892, 65}

\bibitem[\protect\citeauthoryear{{Miranda} \& {Rafikov}}{{Miranda} \&
  {Rafikov}}{2020b}]{miranda20b}
{Miranda} R.,  {Rafikov} R.~R.,  2020b, \mn@doi [\apj]
  {10.3847/1538-4357/abbee7}, \href
  {https://ui.adsabs.harvard.edu/abs/2020ApJ...904..121M} {904, 121}

\bibitem[\protect\citeauthoryear{{M{\"u}ller} et~al.,}{{M{\"u}ller}
  et~al.}{2018}]{Muller2018}
{M{\"u}ller} A.,  et~al., 2018, \mn@doi [\aap] {10.1051/0004-6361/201833584},
  \href {https://ui.adsabs.harvard.edu/abs/2018A&A...617L...2M} {617, L2}

\bibitem[\protect\citeauthoryear{{Nazari}, {Booth}, {Clarke}, {Rosotti},
  {Tazzari}, {Juhasz}  \& {Meru}}{{Nazari} et~al.}{2019}]{Nazari2019}
{Nazari} P.,  {Booth} R.~A.,  {Clarke} C.~J.,  {Rosotti} G.~P.,  {Tazzari} M.,
  {Juhasz} A.,   {Meru} F.,  2019, \mn@doi [\mnras] {10.1093/mnras/stz836},
  \href {https://ui.adsabs.harvard.edu/abs/2019MNRAS.485.5914N} {485, 5914}

\bibitem[\protect\citeauthoryear{{Paardekooper} \& {Mellema}}{{Paardekooper} \&
  {Mellema}}{2006}]{paardekooper06}
{Paardekooper} S.-J.,  {Mellema} G.,  2006, \mn@doi [\aap]
  {10.1051/0004-6361:20054449}, \href
  {http://adsabs.harvard.edu/abs/2006A%26A...453.1129P} {453, 1129}

\bibitem[\protect\citeauthoryear{{Perez}, {Dunhill}, {Casassus}, {Roman},
  {Szul{\'a}gyi}, {Flores}, {Marino}  \& {Montesinos}}{{Perez}
  et~al.}{2015}]{Perez2015b}
{Perez} S.,  {Dunhill} A.,  {Casassus} S.,  {Roman} P.,  {Szul{\'a}gyi} J.,
  {Flores} C.,  {Marino} S.,   {Montesinos} M.,  2015, \mn@doi [\apjl]
  {10.1088/2041-8205/811/1/L5}, \href
  {https://ui.adsabs.harvard.edu/abs/2015ApJ...811L...5P} {811, L5}

\bibitem[\protect\citeauthoryear{{Picogna} \& {Kley}}{{Picogna} \&
  {Kley}}{2015}]{picogna2015}
{Picogna} G.,  {Kley} W.,  2015, \mn@doi [\aap] {10.1051/0004-6361/201526921},
  \href {http://adsabs.harvard.edu/abs/2015A%26A...584A.110P} {584, A110}

\bibitem[\protect\citeauthoryear{{Pinte}, {Dent}, {M{\'e}nard}, {Hales},
  {Hill}, {Cortes}  \& {de Gregorio-Monsalvo}}{{Pinte}
  et~al.}{2016}]{pinte2016}
{Pinte} C.,  {Dent} W.~R.~F.,  {M{\'e}nard} F.,  {Hales} A.,  {Hill} T.,
  {Cortes} P.,   {de Gregorio-Monsalvo} I.,  2016, \mn@doi [\apj]
  {10.3847/0004-637X/816/1/25}, \href
  {http://adsabs.harvard.edu/abs/2016ApJ...816...25P} {816, 25}

\bibitem[\protect\citeauthoryear{{Pinte} et~al.,}{{Pinte}
  et~al.}{2018}]{pinte2018}
{Pinte} C.,  et~al., 2018, \mn@doi [\apjl] {10.3847/2041-8213/aac6dc}, \href
  {http://adsabs.harvard.edu/abs/2018ApJ...860L..13P} {860, L13}

\bibitem[\protect\citeauthoryear{{Rabago} \& {Zhu}}{{Rabago} \&
  {Zhu}}{2021}]{Rabago2021}
{Rabago} I.,  {Zhu} Z.,  2021, \mn@doi [\mnras] {10.1093/mnras/stab447}, \href
  {https://ui.adsabs.harvard.edu/abs/2021MNRAS.tmp..475R} {}

\bibitem[\protect\citeauthoryear{{Rosotti}, {Juhasz}, {Booth}  \&
  {Clarke}}{{Rosotti} et~al.}{2016}]{rosotti2016}
{Rosotti} G.~P.,  {Juhasz} A.,  {Booth} R.~A.,   {Clarke} C.~J.,  2016, \mn@doi
  [\mnras] {10.1093/mnras/stw691}, \href
  {https://ui.adsabs.harvard.edu/abs/2016MNRAS.459.2790R} {459, 2790}

\bibitem[\protect\citeauthoryear{{Rowther}, {Meru}, {Kennedy}, {Nealon}  \&
  {Pinte}}{{Rowther} et~al.}{2020}]{Rowther2020}
{Rowther} S.,  {Meru} F.,  {Kennedy} G.~M.,  {Nealon} R.,   {Pinte} C.,  2020,
  \mn@doi [\apjl] {10.3847/2041-8213/abc704}, \href
  {https://ui.adsabs.harvard.edu/abs/2020ApJ...904L..18R} {904, L18}

\bibitem[\protect\citeauthoryear{{Ruane} et~al.,}{{Ruane}
  et~al.}{2017}]{Ruane2017}
{Ruane} G.,  et~al., 2017, \mn@doi [\aj] {10.3847/1538-3881/aa7b81}, \href
  {https://ui.adsabs.harvard.edu/abs/2017AJ....154...73R} {154, 73}

\bibitem[\protect\citeauthoryear{{Selvaraju}, {Cogswell}, {Das}, {Vedantam},
  {Parikh}  \& {Batra}}{{Selvaraju} et~al.}{2017}]{selvaraju17}
{Selvaraju} R.~R.,  {Cogswell} M.,  {Das} A.,  {Vedantam} R.,  {Parikh} D.,
  {Batra} D.,  2017, in 2017 IEEE International Conference on Computer Vision
  (ICCV). pp 618--626, \mn@doi{10.1109/ICCV.2017.74}

\bibitem[\protect\citeauthoryear{{Simonyan} \& {Zisserman}}{{Simonyan} \&
  {Zisserman}}{2014}]{simonyan14}
{Simonyan} K.,  {Zisserman} A.,  2014, arXiv e-prints, \href
  {https://ui.adsabs.harvard.edu/abs/2014arXiv1409.1556S} {p. arXiv:1409.1556}

\bibitem[\protect\citeauthoryear{{Teague}, {Bae}, {Bergin}, {Birnstiel}  \&
  {Foreman-Mackey}}{{Teague} et~al.}{2018}]{teague18a}
{Teague} R.,  {Bae} J.,  {Bergin} E.~A.,  {Birnstiel} T.,   {Foreman-Mackey}
  D.,  2018, \mn@doi [\apjl] {10.3847/2041-8213/aac6d7}, \href
  {http://adsabs.harvard.edu/abs/2018ApJ...860L..12T} {860, L12}

\bibitem[\protect\citeauthoryear{{Van Der Walt}, {Colbert}  \&
  {Varoquaux}}{{Van Der Walt} et~al.}{2011}]{numpy}
{Van Der Walt} S.,  {Colbert} S.~C.,   {Varoquaux} G.,  2011, preprint, \href
  {http://adsabs.harvard.edu/abs/2011arXiv1102.1523V} {} (\mn@eprint {arXiv}
  {1102.1523})

\bibitem[\protect\citeauthoryear{{Van Oort}, {Xu}, {Offner}  \&
  {Gutermuth}}{{Van Oort} et~al.}{2019}]{vanoort19}
{Van Oort} C.~M.,  {Xu} D.,  {Offner} S. S.~R.,   {Gutermuth} R.~A.,  2019,
  \mn@doi [\apj] {10.3847/1538-4357/ab275e}, \href
  {https://ui.adsabs.harvard.edu/abs/2019ApJ...880...83V} {880, 83}

\bibitem[\protect\citeauthoryear{Virtanen et~al.,}{Virtanen
  et~al.}{2020}]{scipy}
Virtanen P.,  et~al., 2020, \mn@doi [Nature Methods]
  {10.1038/s41592-019-0686-2}, \href {https://rdcu.be/b08Wh} {17, 261}

\bibitem[\protect\citeauthoryear{{Wagner} et~al.,}{{Wagner}
  et~al.}{2018}]{Wagner2018}
{Wagner} K.,  et~al., 2018, \mn@doi [\apjl] {10.3847/2041-8213/aad695}, \href
  {https://ui.adsabs.harvard.edu/abs/2018ApJ...863L...8W} {863, L8}

\bibitem[\protect\citeauthoryear{{Wang} et~al.,}{{Wang}
  et~al.}{2020}]{Wang2020}
{Wang} J.~J.,  et~al., 2020, \mn@doi [\aj] {10.3847/1538-3881/ab8aef}, \href
  {https://ui.adsabs.harvard.edu/abs/2020AJ....159..263W} {159, 263}

\bibitem[\protect\citeauthoryear{Waskom}{Waskom}{2021}]{seaborn}
Waskom M.~L.,  2021, \mn@doi [Journal of Open Source Software]
  {10.21105/joss.03021}, 6, 3021

\bibitem[\protect\citeauthoryear{{Xu}, {Offner}, {Gutermuth}  \& {Oort}}{{Xu}
  et~al.}{2020a}]{xu20a}
{Xu} D.,  {Offner} S. S.~R.,  {Gutermuth} R.,   {Oort} C.~V.,  2020a, \mn@doi
  [\apj] {10.3847/1538-4357/ab6607}, \href
  {https://ui.adsabs.harvard.edu/abs/2020ApJ...890...64X} {890, 64}

\bibitem[\protect\citeauthoryear{{Xu}, {Offner}, {Gutermuth}  \& {Oort}}{{Xu}
  et~al.}{2020b}]{xu20b}
{Xu} D.,  {Offner} S. S.~R.,  {Gutermuth} R.,   {Oort} C.~V.,  2020b, \mn@doi
  [\apj] {10.3847/1538-4357/abc7bf}, \href
  {https://ui.adsabs.harvard.edu/abs/2020ApJ...905..172X} {905, 172}

\bibitem[\protect\citeauthoryear{{Yang} \& {Zhu}}{{Yang} \&
  {Zhu}}{2020}]{yang20}
{Yang} C.-C.,  {Zhu} Z.,  2020, \mn@doi [\mnras] {10.1093/mnras/stz3232}, \href
  {https://ui.adsabs.harvard.edu/abs/2020MNRAS.491.4702Y} {491, 4702}

\bibitem[\protect\citeauthoryear{{Zhang} \& {Zhu}}{{Zhang} \&
  {Zhu}}{2020}]{zhang20}
{Zhang} S.,  {Zhu} Z.,  2020, \mn@doi [\mnras] {10.1093/mnras/staa404}, \href
  {https://ui.adsabs.harvard.edu/abs/2020MNRAS.493.2287Z} {493, 2287}

\bibitem[\protect\citeauthoryear{{Zhang} et~al.,}{{Zhang}
  et~al.}{2018}]{zhang18}
{Zhang} S.,  et~al., 2018, \mn@doi [\apjl] {10.3847/2041-8213/aaf744}, \href
  {http://adsabs.harvard.edu/abs/2018ApJ...869L..47Z} {869, L47}

\bibitem[\protect\citeauthoryear{{Zhou}, {Khosla}, {Lapedriza}, {Oliva}  \&
  {Torralba}}{{Zhou} et~al.}{2016}]{zhou16}
{Zhou} B.,  {Khosla} A.,  {Lapedriza} A.,  {Oliva} A.,   {Torralba} A.,  2016,
  in 2016 IEEE Conference on Computer Vision and Pattern Recognition (CVPR). pp
  2921--2929, \mn@doi{10.1109/CVPR.2016.319}

\bibitem[\protect\citeauthoryear{{Zhu}, {Stone}, {Rafikov}  \& {Bai}}{{Zhu}
  et~al.}{2014}]{zhu14}
{Zhu} Z.,  {Stone} J.~M.,  {Rafikov} R.~R.,   {Bai} X.-n.,  2014, \mn@doi
  [\apj] {10.1088/0004-637X/785/2/122}, \href
  {http://adsabs.harvard.edu/abs/2014ApJ...785..122Z} {785, 122}

\bibitem[\protect\citeauthoryear{{Zhu} et~al.,}{{Zhu} et~al.}{2019}]{zhu19b}
{Zhu} Z.,  et~al., 2019, \mn@doi [\apjl] {10.3847/2041-8213/ab1f8c}, \href
  {https://ui.adsabs.harvard.edu/abs/2019ApJ...877L..18Z} {877, L18}

\bibitem[\protect\citeauthoryear{{Ziampras}, {Kley}  \& {Dullemond}}{{Ziampras}
  et~al.}{2020}]{ziampras20}
{Ziampras} A.,  {Kley} W.,   {Dullemond} C.~P.,  2020, \mn@doi [\aap]
  {10.1051/0004-6361/201937048}, \href
  {https://ui.adsabs.harvard.edu/abs/2020A&A...637A..50Z} {637, A50}

\bibitem[\protect\citeauthoryear{{Zurlo} et~al.,}{{Zurlo}
  et~al.}{2020}]{Zurlo2020}
{Zurlo} A.,  et~al., 2020, \mn@doi [\aap] {10.1051/0004-6361/201936891}, \href
  {https://ui.adsabs.harvard.edu/abs/2020A&A...633A.119Z} {633, A119}

\makeatother
\end{thebibliography}







\bsp	
\label{lastpage}
\end{document}